\documentclass[fleqn,usenatbib]{mnras}

\usepackage{newtxtext,newtxmath}
\usepackage[T1]{fontenc}
\usepackage{ae,aecompl}
\usepackage{graphicx}	
\usepackage{amsmath}
\newcommand{\vsini}{\ensuremath{{\upsilon}\sin i}}
\newcommand{\kms}{\,km\,s$^{-1}$}
\newcommand{\zav}[1]{\left(#1\right)}
\newcommand{\hzav}[1]{\left[#1\right]}

\newcommand{\hd}{HD\,174356}
\newcommand{\Teff}{$T_\mathrm{eff}$}
\newcommand{\lam}{$\lambda$}

\title[Single g-mode pulsation in the B9pSi star HD\,174356?]{Rotational modulation and single g-mode pulsation in the B9pSi star HD\,174356?}

\author[Z.~Mikul{\'a}{\v s}ek et al.]{Z.~Mikul{\'a}{\v s}ek,$^{1}$\thanks{E-mail: mikulas@physics.muni.cz}
   E.~Paunzen,$^{1}$
   S.~H{\"u}mmerich,$^{2,3}$
	 E.~Niemczura,$^{4}$
	 P.~Walczak,$^{4}$
	 L.~Fraga,$^{5}$
	\newauthor
	 K.~Bernhard,$^{2,3}$
	 J.~Jan\'{i}k,$^{1}$
     S.~Hubrig,$^{6}$
     S.~J{\"a}rvinen,$^{6}$
	 M.~Jagelka,$^{1}$
	 O.~I.~Pintado,$^{7}$
	\newauthor
   J.~Krti\v{c}ka,$^{1}$
   M.~Pri{\v s}egen,$^{1}$	
   M.~Skarka,$^{1,8}$
   M.~Zejda,$^{1}$
   I.~Ilyin,$^{6}$
   T.~Pribulla,$^{9,10}$
	 K.~Kami\'{n}ski,$^{11}$
	\newauthor
   M.~K.~Kami\'{n}ska,$^{11}$
   J.~Tokarek,$^{11}$
   and P.~Zieli\'{n}ski$^{12}$
	\\
$^{1}$Department of Theoretical Physics and Astrophysics, Masaryk University, Kotl{\'a}\v{r}sk{\'a} 2, CZ\,611\,37, Czech Republic\\
$^{2}$Bundesdeutsche Arbeitsgemeinschaft f{\"u}r Ver{\"a}nderliche Sterne e.V. (BAV), Berlin, Germany\\
$^{3}$American Association of Variable Star Observers (AAVSO), Cambridge, USA\\
$^{4}$Astronomical Institute, University of Wroc{\l}aw, Wroc{\l}aw, Poland\\
$^{5}$Laborat{\'o}rio Nacional de Astrof{\'i}sica LNA/MCTIC, Itajub{\'a}, Brazil\\
$^{6}$Leibniz-Institut f{\"u}r Astrophysik Potsdam (AIP), Potsdam, Germany\\
$^{7}$Centro de Tecnolog{\'i}a Disruptiva, Universidad de San Pablo Tucum{\'a}n, San Pablo, Tucum{\'a}n, Argentina\\
$^{8}$Astronomical Institute, Czech Academy of Science, Czech Republic\\
$^{9}$Astronomical Institute of the Slovak Academy of Science, Tatransk{\'a} Lomnica, Slovakia\\
$^{10}$ELTE Gothard Astrophysical Observatory, 9700 Szombathely, Szent Imre h. u. 112, Hungary\\
$^{11}$Institute Astronomical Observatory, Faculty of Physics, Adam Mickiewicz University, Pozna\'{n}, Poland\\
$^{12}$Astronomical Observatory University of Warsaw, Warsaw, Poland\\
}
	
\date{Accepted XXX. Received YYY; in original form ZZZ}
\pubyear{2020}

\begin{document}
\label{firstpage}
\pagerange{\pageref{firstpage}--\pageref{lastpage}}
\maketitle

\begin{abstract}
Chemically peculiar (CP) stars of the upper main sequence are characterised by specific anomalies in the photospheric abundances of some chemical elements. The group of CP2 stars, which encompasses classical Ap and Bp stars, exhibits strictly periodic light, spectral, and spectropolarimetric variations that can be adequately explained by the model of a rigidly rotating star with persistent surface structures and a stable global magnetic field. Using observations from the {\it Kepler} K2 mission, we find that the B9pSi star HD\,174356 displays a light curve both variable in amplitude and shape, which is not expected in a CP2 star. Employing archival and new photometric and spectroscopic observations, we carry out a detailed abundance analysis of HD\,174356 and discuss its photometric and astrophysical properties in detail. We employ phenomenological modeling to decompose the light curve and the observed radial velocity variability. Our abundance analysis confirms that HD\,174356 is a silicon-type CP2 star. No magnetic field stronger than 110\,G was found. The star's light curve can be interpreted as the sum of two independent strictly periodic signals with $P_1=4\fd043\,55(5)$ and $P_2=2\fd111\,69(3)$. The periods have remained stable over 17 years of observations. In all spectra, HD\,174356 appears to be single-lined. From the simulation of the variability characteristics and investigation of stars in the close angular vicinity, we put forth the hypothesis that the peculiar light variability of HD\,174356 arises in a single star and is caused by rotational modulation due to surface abundance patches ($P_1$) and g mode pulsation ($P_2$).
\end{abstract}


\section{Introduction} \label{introduction}

Chemically peculiar (CP) stars constitute about 10-15\,\% of the upper main-sequence stars between spectral types B2 and F5 \citep{preston74}.
They are characterised by abundance anomalies of individual chemical elements (typically Si, He, Fe, rare earths, Co, Ni, Ti, Sr, Sc, Ca, and others) in their outer layers that may reach several orders of magnitude as compared to the solar values. The observed peculiar abundances are generally explained by chemical segregation due to the interplay of radiative levitation and gravitational settling taking place in the calm radiative outer layers of mostly slowly rotating stars \citep[e.g.][]{richer00}.

\begin{figure*}
\begin{center}
\includegraphics[width=0.8\textwidth]{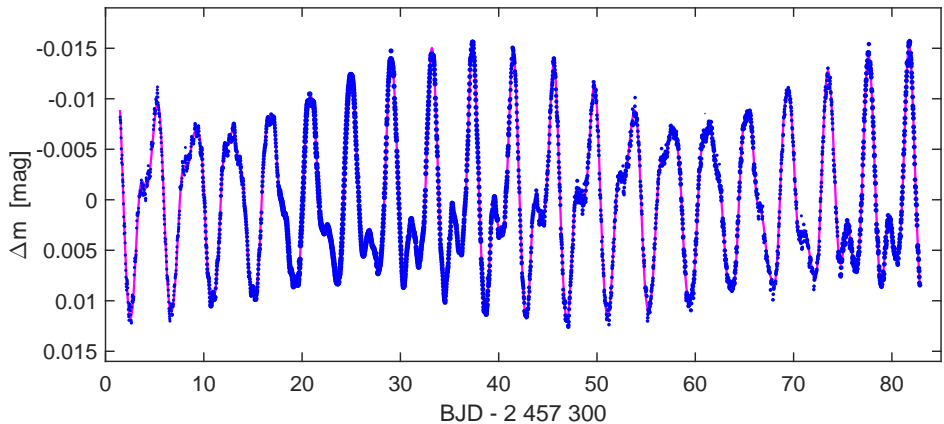}
\caption{Detrended light curve of \hd\ based on {\it Kepler}\,K2 observations from the MAST archive (blue dots). The model, which is approximated by the linear combination of two independent strictly periodic functions (see Eq.\,(\ref{model})) with periods of 4.043\,61(5)\,d and 2.111\,65(3)\,d (Table\,\ref{tab}), is shown in pink and describes the observed K2 light curve with impressive accuracy.}
\label{detrlc}
\end{center}
\end{figure*}

Chemically peculiar stars are divided into several types according to their abundance patterns and effective temperatures. The group of CP2 stars, which encompasses classical Ap and Bp stars, has been well studied. Together with the He-weak and He-strong stars (also termed CP4/5 and CP6/7 stars; cf. \citealt{maitzen84}), these objects are also referred to as magnetic chemically peculiar (mCP) stars because of the near-ubiquitous presence of a globally-organised magnetic field that may attain strengths of up to several tens of kG \citep{babcock58,babcock60,mathys01}. We note, however, that the mCP classification might be occasionally misleading, as we know several well-studied CP2 stars with very weak \citep[$\varepsilon$ UMa;][]{shulyak10} or unmeasurable \citep[EE Dra;][]{krticka09} magnetic fields.

Besides the detection of a magnetic field, there exist other indicators of mCP stars, such as the presence of strictly periodic spectral, photometric, and spectropolarimetric variability that is well described by the oblique rotator model \citep{stibbs50}. The spectral changes indicate an uneven surface distribution of overabundant/underabundant chemical elements (abundance spots and patches) that persists for decades and more. The photometric variations, traditionally referred to as $\alpha^2$ Canum Venaticorum (ACV) variability \citep{gcvs}, are caused by the redistribution of flux in the abundance patches \citep[e.g.][]{krticka07,krticka13}, while the periodic spectropolarimetric changes are the result of rotation in the presence of a more or less dipole/quadrupole-like magnetic field that is frozen into the outer layers of the star. A direct link between magnetic field geometry and the location of the spectroscopic and photometric spots on mCP stars has not been proven \citep{2019A&A...622A.199J}. Nevertheless, we assume that, in addition to calm photospheres and slow rotation, global magnetic fields contribute to the stabilisation of the surface structures.

In the spectral realm of the mCP stars, we also encounter classical pulsators \citep{2014IAUS..301...31P}. Pulsating stars allow to probe stellar interiors that are not accessible to direct observations because the physical conditions and processes within a star significantly influence the observed pulsation periods and amplitudes. In this way, basic stellar properties such as mass, radius and distance can be derived \citep{2011A&A...530A.142B,2015AN....336..477A}.

\section{\hd\ -- basic information}\label{targetstar}

\hd\ (CD $-$24\,14762, TYC 6864-121-1) was identified as a standard CP2 star of type B9pSi in the past \citep{houk88,rm09}. However, the star was found to display a light curve both variable in amplitude and shape (see Fig.\,\ref{detrlc}), which is not expected in a CP2 star.

The light variability of \hd\ was first reported by \citet{huemmerich16}, who listed a rotational period of 4.0431(3)\,d based on ASAS-3 photometry. Recently, \citet{bowman18} confirmed this period using {\it Kepler}\,K2 data. They furthermore detected a second period in the amplitude spectrum, which is not well-resolved from the harmonic of the measured rotation period. They suggested that \hd\ may be a binary or multiple system and pointed out the similarity between its light curve and the light curves of cool differentially-rotating spotted stars. No further analysis of the variability characteristics was done in \citet{bowman18}.

The Data Release 2 of the $Gaia$ satellite mission \citep{brown16,gaiacol18,2018A&A...616A...2L}\footnote{http://vizier.u-strasbg.fr/viz-bin/VizieR-3?-source=I/345/gaia2} lists a parallax of $\pi$\,=\,1.17(11)\,mas, which converts to a distance modulus $\mu=9.66(21)$\,mag. Employing the 3-D dust maps of the Pan-STARRS1 project \citep{green15}, we find that the star is located in a region of the Milky Way ($l$\,=\,10\fdg93 and $b$\,=\,$-$10\fdg79) significantly affected by interstellar extinction. For the above listed distance modulus, a reddening $E({\it B-V})=0.30(2)$\,mag was deduced from the maps.

Adopting these values, the observed ($B-V$ index of 0.21\,mag, corrected for interstellar reddening of {$(B-V)_0=-0.09(2)$\,mag, locates the object in the region of the late B-type stars \citep[\Teff\,$\approx$\,12\,000\,K,][]{kharchenko01}, which is consistent with the observed 2MASS $JHK_\mathrm{S}$ \citep{2mass} photometry. Assuming $A_V/E(B-V)\cong3.1$ \citep[e.g.][]{majaess16,berdnikov96,pejcha12}, we estimate an apparent magnitude of $V_{0}=8.23(7)$\,mag and an absolute magnitude of $M_V=-1.43(22)$\,mag, both corrected for dust extinction. These values also agree with the observed spectral type and the derived effective temperature.

In this paper, we study and discuss the photometric and spectroscopic properties of \hd, with the goal of revealing the cause of its unique light variability and the nature of the object(s) involved. To this end, we employ and analyse archival observations and own data of various kinds.

\section{Observations}

In the following sections, we provide information on the photometric and spectroscopic observations used in this study.

\subsection{Photometry} \label{section_photometry}

Three different data sources and own observations were employed, which are described in more detail below. \hd\ is not scheduled for any TESS observations.\footnote{https://heasarc.gsfc.nasa.gov/cgi-bin/tess/webtess/wtv.py} The characteristics of the employed photometric data are summarized in Table\,\ref{photom}.

\subsubsection{ASAS-3 observations} \label{asas3} %done

The All Sky Automated Survey (ASAS) monitored the southern and part of the northern sky (up to $\delta$\,$<$\,+28\degr), with the goal of investigating any kind of stellar photometric variability. Here we employ Johnson $V$ observations from the third phase of the project, ASAS-3, which were obtained with two wide-field telescopes equipped with f/2.8 200\,mm Minolta lenses and $2048\times2048$ AP 10 Apogee detectors (sky coverage $9^{\circ}\!\times9^{\circ}$) located at the ten-inch astrograph dome of the Las Campanas Observatory in Chile. The typical exposure time was three minutes. The most reliable light curves were obtained for objects in the magnitude range 8\,$\le$\,$V$\,$\le$\,10\,mag (typical scatter of about 0.01\,mag; \citealt{Pigulski14}). Frequency identification is rendered difficult because of strong daily aliases in the Fourier spectra based on ASAS-3 data. In this work, we employ only data points with quality flags ``A'' and ``B'', which identify the most reliable data. Obvious outliers were removed by visual inspection.

\begin{table}
\begin{center}
\caption{Characteristics of the employed photometric data. The columns denote, respectively, the data source, the covered time span, the number of observations available, and the scatter.}\label{photom}
\begin{tabular}{lclrc}
  \hline
Source & Filter & Coverage & $N_{\rm {obs}}$ & $s$   \\
& & & & [mmag] \\
 \hline
 ASAS-3 & $V$ & 2001-2009 & 590& 12\\
 {\it Kepler}\,K2 & &2015-2016 & 3718 & 0.5\\
 ASAS-SN & $V$ & 2016-2018 &433& 23\\
\hline
\end{tabular}
\end{center}
\end{table}

\subsubsection{{\it Kepler}\,K2 observations}\label{kepler_k2} %done

The $Kepler$ spacecraft boasts a differential photometer with 0.95\,m aperture and detectors consisting of 21 modules each equipped with $2200\times1024$ pixel CCD detectors. It produces ultra-precise single-passband (420-900\,nm) light curves obtained in long-cadence (29.4\,min) and short-cadence (58.5\,s) modes. After the loss of two reaction wheels, the $Kepler$ spacecraft was recommissioned to observe the ecliptic plane in the K2 mission \citep{howell14}. To adjust the pointing of the spacecraft, the on-board thrusters are fired at intervals of about six hours, which leads to characteristic systematics in the data sets \citep{howell14}.

About 3700 measurements in long-cadence mode were obtained of \hd\ during a time span of 81.3 days (October 4 to December 26, 2015) in Campaign 7 of the K2 mission. The corresponding light curve was procured from the archive of K2 Data Products at the Mikulski Archive for Space Telescopes.\footnote{https://archive.stsci.edu/k2/}

The thruster firing signature at integers of $\sim$4.08\,d$^{-1}$ is obvious in the data. It is, however, of comparably low amplitude and does not interfere with the frequency analysis. We investigated the plots of \citet{luger16} to check for contamination of the light curve by neighbouring sources; no significant contaminating sources were identified.

\subsubsection{ASAS-SN observations} \label{asas_sn} %done

All-Sky Automated Survey for Supernovae (ASAS-SN) observations are obtained at five stations, each of which is equipped with four 14 cm aperture Nikon telephoto lenses. Every clear night, the entire visible sky is observed to a depth of $V$\,$<$\,17\,mag in three dithered 90\,s exposures made through $V$ or $g$ band filters \citep{Shappee2014,Kochanek2017}. The available data boast time baselines of up to five years. Depending on camera and image position, saturation issues set in at 10 to 11\,mag. The effects of saturation are counterbalanced using a procedure inherited from the original ASAS survey, which is described in detail in \citet{jayasinghe18}.

\subsubsection{Own observations of \hd\ and stars in its vicinity}\label{CCD}

We have carried out CCD photometric observations of \hd\ and its vicinity using the STE4 camera attached to the 1\,m telescope on the SAAO at Sutherland, South Africa (March 2018), and the 1.54\,m Danish telescope at the La Silla Observatory in Chile (April 2018). At both sites, a Johnson $B$ filter was used. The seeing was between 0.8\arcsec\ and 1.2\arcsec.

\subsection{Spectroscopy and spectropolarimetry}\label{Spectroscopy}

To determine astrophysical parameters, investigate the presence of chemical peculiarities and magnetic field indicators, and look for possible signs of binarity, as suggested by \citet{bowman18}, \hd\ was investigated spectroscopically. Spectra from the following instruments were used:
\begin{itemize}
\item EBASIM Spectrograph at the 2.15m telescope at Complejo Astron{\'o}mico El Leoncito (CASLEO, Argentina), 226 lines/mm grating, R\,$\sim$\,40\,000,
4520\,\AA\ to 6520\,\AA, September 2018.
\item Fiber-fed Extended Range Optical Spectrograph (FEROS) spectrograph attached to the 2.2m MPG/ESO telescope at La Silla Observatory (ESO) in Chile, $R=48\,000$, 3705\,\AA\ to 9225\,\AA, August 2017, Prog. ID: 099.A-9039(C).
\item Goodman High Throughput Spectrograph (GTHS) at the 4.1m Southern Astrophysical Research (SOAR) telescope, red camera, 2100 lines/mm grating, R\,$\sim$\,11\,000, 4460\,\AA\ to 5090\,\AA, September 2018, April 2019.
\item High Accuracy Radial velocity Planet Searcher (HARPS) spectrograph at the ESO La Silla 3.6m telescope in Chile, R\,$\sim$\,110\,000, 3900\,\AA\ to 6900\,\AA, June 2017, Prog. ID: 099.C-0081(A).
\item HIgh-Dispersion Echelle Spectrograph (HIDES) at the 1.88m telescope of the Okayama Astrophysical Observatory (OAO) in Japan, R\,$\sim$\,50\,000, 4090\,\AA\ to 7\,520\,\AA, May 2017.
\item Pozna\'n Spectroscopic Telescope~2 (PST2) located at Winer Observatory in Arizona, R\,$\sim$\,40\,000, 3890\,\AA\ to 9130\,\AA, March 2017.
\end{itemize}
Spectra were bias-corrected and flat-fielded, and data were reduced using standard IRAF routines.\footnote{IRAF is distributed by NOAO, which is operated by AURA, Inc., under cooperative agreement with the National Science Foundation.} A Th-Ar-Ne comparison spectrum was employed for precise wavelength calibration of the FEROS spectra, which was simultaneously obtained through a second fiber. In the case of the other instruments, we used Th-Ar spectra that were measured before and after the stellar spectrum.

The spectropolarimetric observations obtained with HARPSpol were reduced and calibrated with the HARPS data reduction software that is available at ESO. The normalization of the spectra to the continuum level was described in detail by \citet{hubrig13}.

\section{\emph{Analysis}}

\subsection{Atmospheric parameters} \label{atmo} %done

Following \citet{Catanzaro2010}, we determined effective temperature \Teff\ and surface gravity $\log g$ by comparing the observed and synthetic hydrogen H$\beta$ and H$\alpha$ lines using the averaged FEROS spectrum, which does not suffer from normalization issues of the strongly broadened hydrogen lines. The corresponding uncertainties were calculated by taking into account the differences in the values obtained from the individual Balmer lines.

In a subsequent step, the \Teff\ and $\log g$ values obtained from the hydrogen-line fitting process were checked by analysis of the \ion{Fe}{I} and \ion{Fe}{II} lines. We here used the high-resolution averaged HARPS spectrum and adjusted the effective temperature until no trend in the abundance versus excitation potential for the \ion{Fe}{II} lines remained. To determine $\log g$, we relied on the ionization equilibrium of the \ion{Fe}{I} and \ion{Fe}{II} lines. Furthermore, microturbulence $\xi$ was adjusted in such a way that trends of \ion{Fe}{II} abundance versus line strength were removed. Simultaneously, we obtained the projected rotational velocity $v\sin i$. The employed spectrum synthesis method allows the simultaneous determination of various interlinked parameters such as \Teff, $\log g$, $\xi$, $v\sin i$, and the relative abundances of the elements. Atmospheric parameters were obtained before the chemical abundance analysis.

We used one-dimensional plane-parallel hydrostatic models that assume radiative equilibrium and were calculated with the ATLAS\,9 code \citep{Kurucz2014}. Synthetic spectra were obtained using the line-blanketed local thermodynamical equilibrium code SYNTHE \citep{Kurucz2005}. The codes were ported to GNU/Linux by \citet{Sbordone2005}. We made use of the most recent line list available at the Fiorella Castelli website.\footnote{http://wwwuser.oats.inaf.it/castelli/}

\begin{table}\scriptsize
\caption{Elemental abundances ($\log N/N_{\rm{tot}}+12$) of \hd. The columns denote, respectively, the corresponding element (and ordinal number), the number of lines analyzed, the derived abundance value (and uncertainty), and the solar abundance value (SA) from \citet{Asplund2009}. Overabundant and underabundant elements are emphasised by the use of bold and italic fonts.}
\label{table_abund}
\centering
\begin{tabular}{lrcr|lrcr}
\hline\hline
Elem. & $N$ & Abund. & SA & Elem. & $N$ & Abund. & SA \\
\hline
\emph{He} (2)  & 4 & 10.19(20) & 10.93  &\textbf{Ti} (22) & 8 & 5.43(24) & 4.95   \\
C (6)   & 4 & 8.69(25) & 8.43    & V (23)  & 2 & 3.93(1) & 3.93 \\
N (7)   & 2 & 7.31(1) & 7.83     & \textbf{Cr} (24) & 22 & 6.15(30) & 5.64 \\
O (8)   & 4 & 8.59(25) & 8.69    & \textbf{Mn} (25) & 13 & 6.63(27) & 5.43 \\
Ne (10) & 6 & 8.30(7) & 7.93     & \textbf{Fe} (26) & 102 & 7.85(15) & 7.50 \\
Na (11) & 2 & 7.78(1) & 6.24     & \textbf{Co} (27) & 1 & 6.20 & 4.99 \\
\emph{Mg} (12) & 3 & 6.90(20) & 7.60    & Ni (28) & 8 & 6.20(28) & 6.22 \\
Al (13) & 4 & 5.68(10) & 6.45    & \textbf{Sr} (38) & 1 & 4.16 & 2.87 \\
\textbf{Si} (14) & 21 & 8.34(20) & 7.51 & \textbf{Y} (39)  & 1 & 4.10 & 2.21 \\
\textbf{P }(15)  & 21 & 7.25(16) & 5.41   & \textbf{Zr} (40) & 1 & 3.89 & 2.58 \\
S (16)  & 17 & 6.93(31) & 7.12   & \textbf{Xe} (54) & 3 & 6.60(20) & 2.24 \\
\textbf{Ar} (18) & 2 & 7.42(1) & 6.40     & \textbf{Pr} (59) & 2 & 3.73(1) & 0.72 \\
Ca (20) & 1 & 6.62 & 6.34        & \textbf{Nd} (60) & 3 & 3.83(9) & 1.42 \\
\hline
\end{tabular}
\end{table}

The following parameters were determined: \Teff$=13\,200$\,K, $\log g = 3.8$(2), $\xi = 0.5(2)$\,\kms, and $v\sin i = 40(3)$\,\kms. The derived effective temperature agrees well with the star's $BV$~$JHK_\mathrm{S}$ colours (Sect.\,\ref{targetstar}) and fits the published spectral type of B9pSi very well.

Table \ref{table_abund} compares the derived elemental abundances to the solar abundance values from \citet{Asplund2009}. The abundance pattern of our target star clearly resembles that of classical CP2 stars \citep{Lopez2001}. In particular, silicon, the iron-peak elements, and the rare-earths are strongly enhanced, while helium is considerably deficient. Helium deficiency is symptomatic not only for the He-weak but also for the cooler mCP B-type stars.

The fit to the observed spectrum in selected spectral regions is shown in Fig.~\ref{lines}. We here note that the employed FEROS and HARPS spectra appear single-lined.

\subsection{Search for a longitudinal magnetic field}\label{spepol} %done

High-resolution spectropolarimetric observations of \hd\ were obtained with the HARPSpol instrument on three consecutive nights in 2017 June 4--6 (cf. Section \ref{Spectroscopy}, the observation log is listed in Table \ref{T:Bz}). To study the presence of a mean longitudinal magnetic field, we employed the Least-Square Deconvolution (LSD) technique \citep{donati97} as described in \citet{2018A&A...618L...2J}. We checked that the lines selected for the line mask are indeed visible in the spectra. The mean longitudinal magnetic field was evaluated by computing the first-order moment of the Stokes $V$ profile according to \citet{mathys89}.

\begin{figure}
\begin{center}
\includegraphics[width=0.45\textwidth]{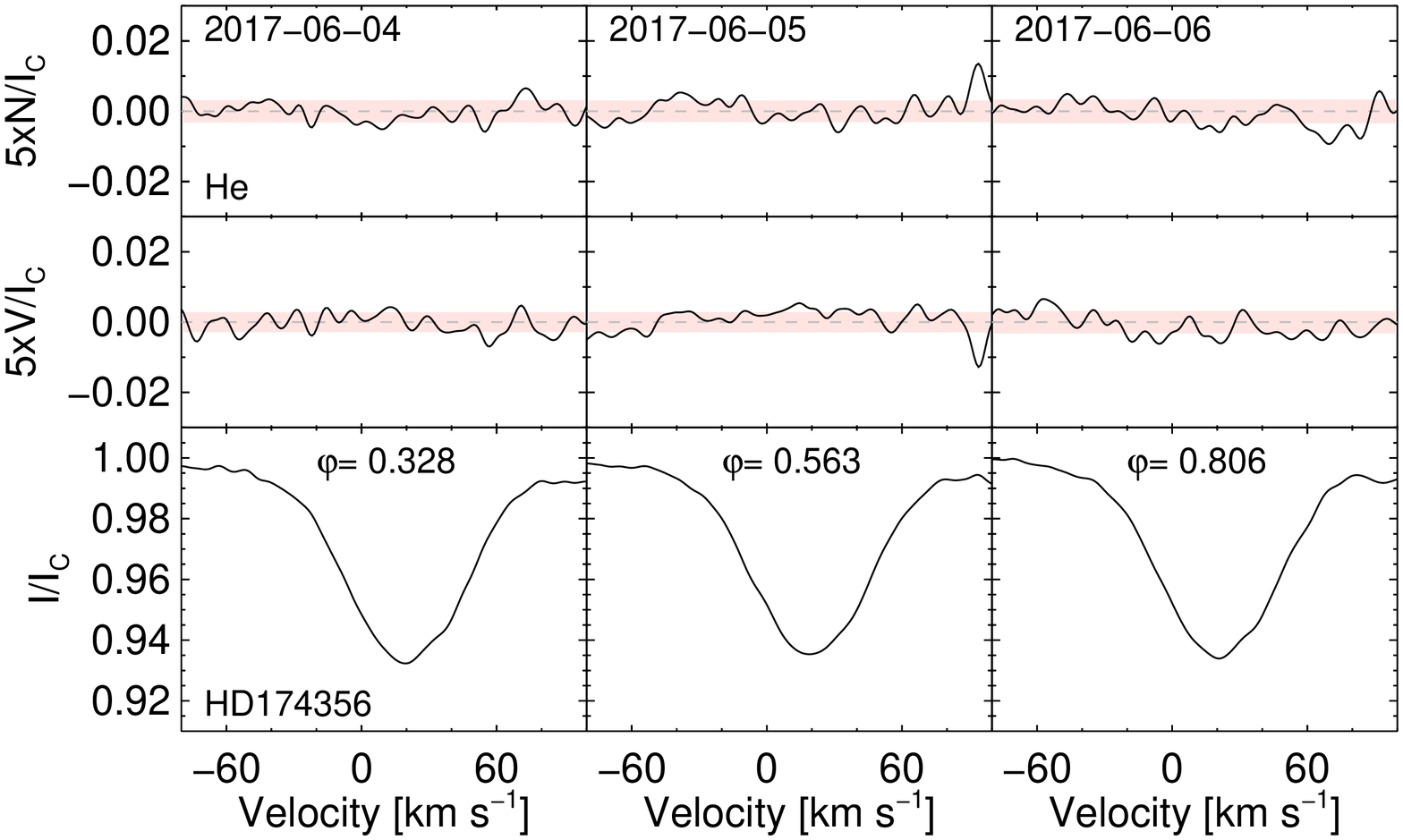}
\includegraphics[width=0.45\textwidth]{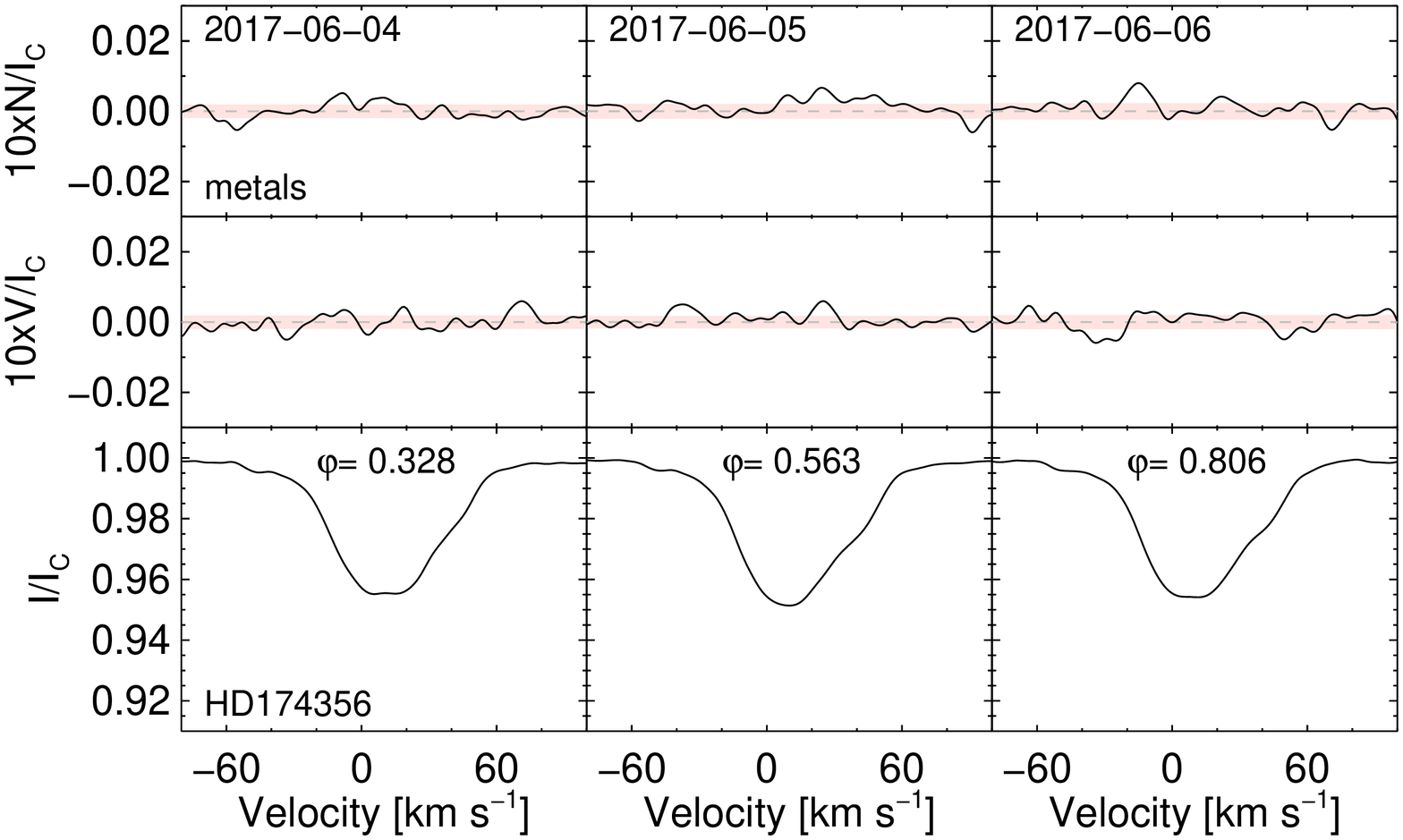}
\caption{LSD Stokes I, V, and diagnostic null spectrum $N$ profiles calculated for
different nights using a line mask containing exclusively He lines
(upper panel) and a line mask with metal lines (lower panel).
Shaded regions indicate the mean uncertainty. In addition to the dates
of the observations, the related rotational phases are
indicated in the plots.
}
\label{Stokes}
\end{center}
\end{figure}

To take into account the effect of the inhomogeneous surface element distribution usually observed in mCP stars, we decided to use two different line masks for the measurements - one including all metal lines apart from the hydrogen and helium lines, and one mask containing He lines exclusively. The He-line mask has 25 lines; the metal (Fe, Si, Cr, C, and Ti) mask has 111 lines. The resulting LSD Stokes $I$, $V$, and diagnostic $N$ profiles are illustrated in Fig.~\ref{Stokes}.

As is evident from Fig.~\ref{Stokes}, no magnetic field is detected in HARPSpol data. For almost all cases, the false alarm probability (FAP) is larger than $10^{-3}$, which is usually considered the limit for detection. FAP is commonly used to classify the magnetic field detection with the LSD technique. The limits for non-detections and detections were originally introduced by \citet{donati97}. Only the LSD Stokes $V$ profile obtained on 2017 June 6 with the mask containing metal lines presents a feature similar to a crossover  signature indicating a marginal detection of a longitudinal magnetic field of $-70 \pm 40$ G with FAP smaller than $10^{-3}$. Only changes in the He line profiles are visible. Because He is usually concentrated at the magnetic poles, it is possible that we observe the star in the direction of the magnetic equator. The detected weak crossover signature possibly indicates that this star possesses a weak magnetic field. Based on our measurements, we conclude that the mean longitudinal magnetic field of HD 174356 cannot exceed 110\,G.

It has been suggested that there exists a dichotomy between strong and ultra-weak magnetic fields among intermediate-mass stars \citep{2014IAUS..302..338L}. The so-called ``magnetic desert'' ranges from about 100\,G to a few G \citep{2013MNRAS.428.2789B}. \hd\ may be located in just this region.

\subsection{Photometric variability}\label{Photometry} %done

We have at our disposal 4741 individual measurements of \hd\ that have been obtained during a time span of 17 years and procured from three different sources (see Section \ref{section_photometry} and Table\,\ref{photom}).

\subsubsection{Amplitude frequency spectra} %done

The detrended light curve of \hd\ shows peculiar light variability with a fundamental period of 4.04\,d that is susceptible to variations in amplitude and shape on a time scale of about 48\,days (see Fig.\,\ref{detrlc}). This behavior can be traced not only in K2 data but in all photometric data that we used.

The key to comprehend this type of light variability and make the best use of the available observations is provided by weighted frequency analysis, which enables to study dependencies of the amplitudes of periodic variations versus their frequencies for data of uneven quality. To this end, periodograms as described in \citet{mikper15} were employed, which also allow to assess the successfulness of the phenomenological modeling of the observations and predict the positions of aliases caused by data sampling and light curve harmonics.

\begin{figure}
\begin{center}
\includegraphics[width=0.5\textwidth]{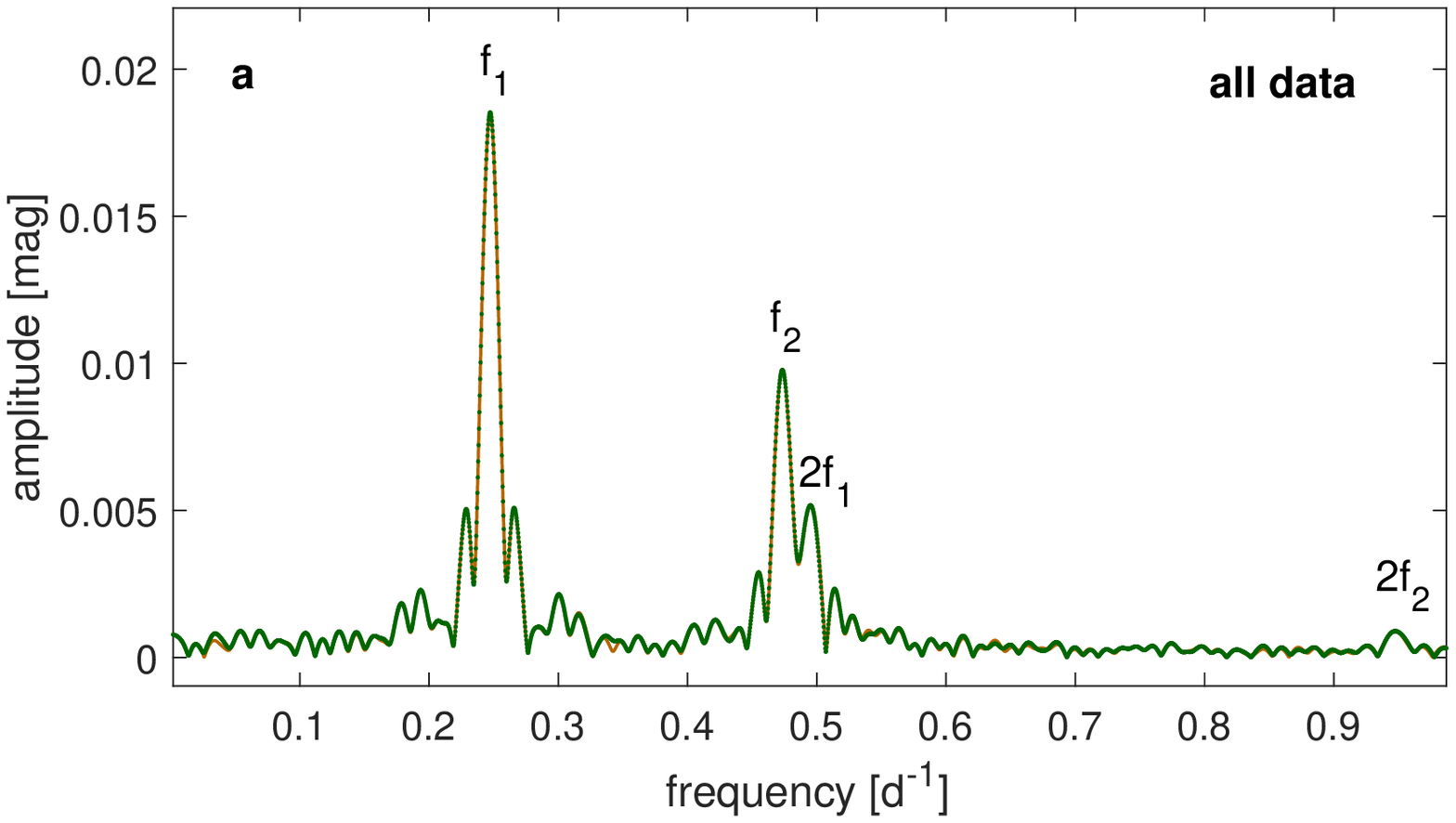}
\includegraphics[width=0.5\textwidth]{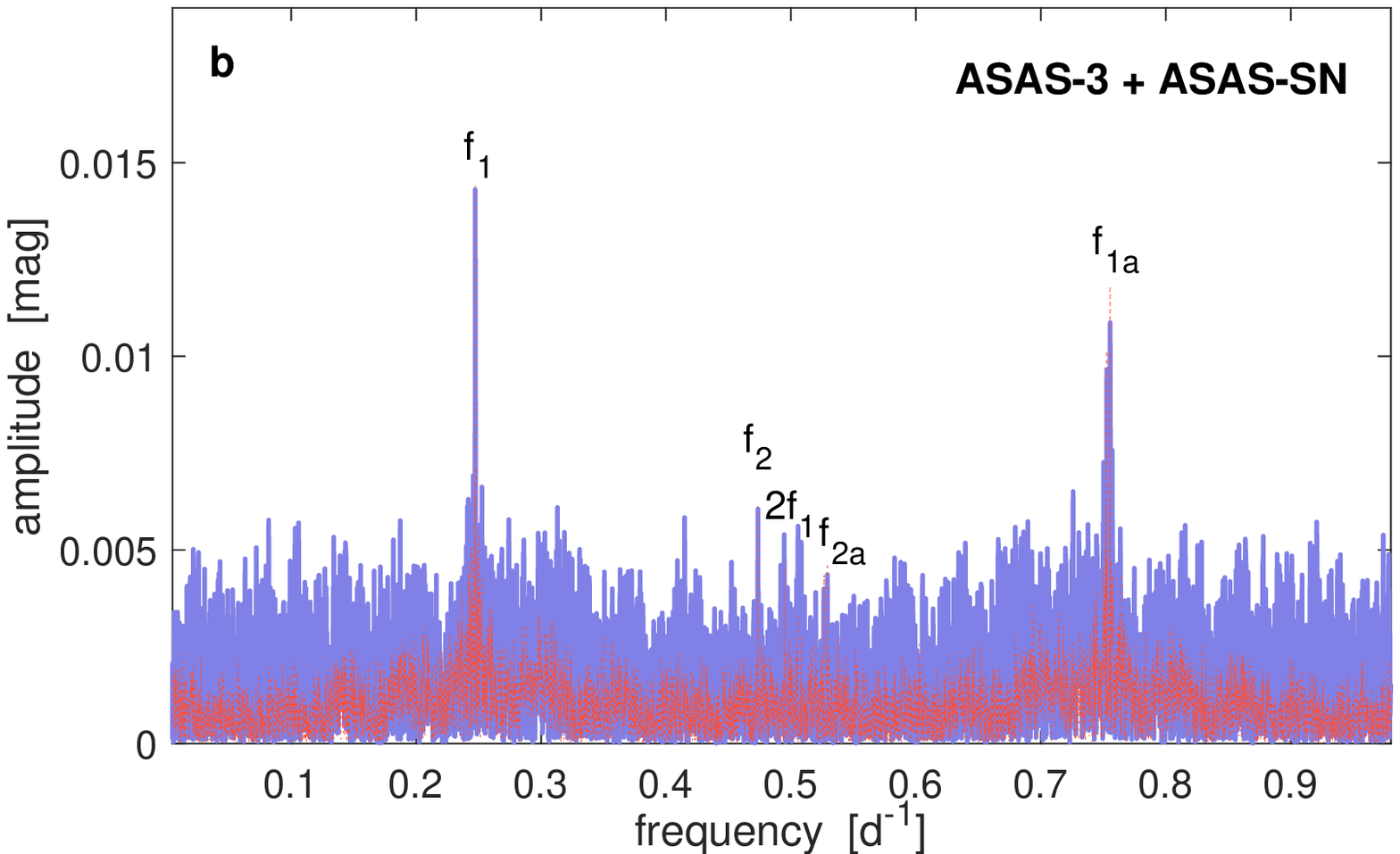}
\caption{Amplitude frequency spectra of \hd, created from photometric data and the corresponding models (Eq.\,{\ref{model}}). Panel (a) illustrates the spectrum for all weighted data (green dots) and the corresponding model (brown lines). Panel (b) shows the spectrum for only ASAS-3 and ASAS-SN data (lilac lines) and the corresponding model (red dots). The light curve components with frequencies $f_1$ and $f_2$ are well visible in both spectra. The ground-based data are also indicative of the one-day alias frequencies $f_{1\rm a}$ and $f_{2\rm a}$.}
\label{periodog}
\end{center}
\end{figure}

\begin{figure}
\begin{center}
\includegraphics[width=0.53\textwidth]{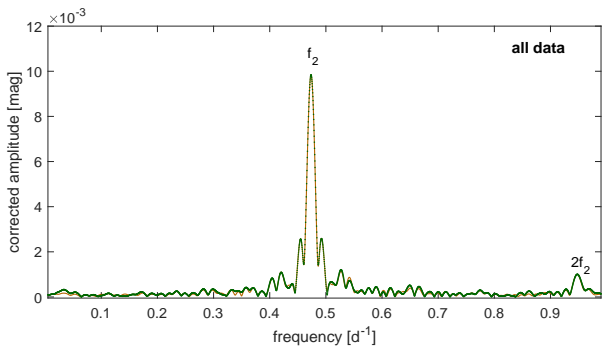}
\caption{Amplitude frequency spectrum created from all weighted data after subtracting the frequency component $f_1$ (green dots) and the corresponding model (brown lines). The resulting spectrum is extraordinarily pure and dominated by the second frequency component, which seems to be totally independent of the first one.}
\label{periodog2}
\end{center}
\end{figure}

Panel (a) of Fig.\,\ref{periodog} depicts the amplitude frequency spectrum created from all photometric data, weighted inversely to the square of their accuracy (see Table\,\ref{photom}). In consequence, the periodogram is dominated by the K2 observations that were obtained continuously but only for a time span of 81 days. In the frequency range from 0 to 1 d$^{-1}$, the periodogram pinpoints the two principle independent frequencies $f_1=0.2473$\,d$^{-1}$ and $f_2=0.4736$\,d$^{-1}$. The other conspicuous peaks correspond to the first harmonics of these frequencies $2\,f_1=0.4946$\,d$^{-1}$ and $2\,f_2=0.9471$\,d$^{-1}$. Two subsidiary peaks related to the time interval of the K2 observations are symmetrically placed around the main peaks. The amplitude changes visible in Fig.\,\ref{detrlc} take place on a period of $\sim$48\,d and follow the relation $1/(2f_1-f_2)$.

Panel (b) of Fig.\,\ref{periodog} shows the amplitude frequency spectrum created from ground-based ASAS observations collected over a time interval of 17 years. The dominant peaks are centered at the basic frequencies $f_1$ and $f_2$. Also visible are the corresponding one sidereal day aliases. The periodogram shows that both frequencies are present also in ASAS-3 and ASAS-SN data. We have also tried to find oscillations in the high frequency range. Several peaks with amplitudes of up to 1\,mmag are present in the 3.6 to 4.6\,d$^{-1}$ interval. However, careful inspection shows that these are artifacts -- aliases of the two main components caused by the uneven distribution of observations within the K2 season.

After subtracting the frequency component $f_1$, the resulting amplitude frequency spectrum is extraordinarily pure (see Fig.\,\ref{periodog2}), without any signs of corruption introduced by the removal of the main frequency. Thus, the light variations in both frequencies are entirely independent and might stem from two independent sources. This conclusion is the starting point for the construction of our general two-component phenomenological model of the observed light curve, which is discussed in Sect. \ref{phenmodel}.

\subsubsection{Probing the variability of stars in the close angular vicinity} \label{vicinity}

\begin{figure}
\begin{center}
\includegraphics[width=0.20\textwidth]{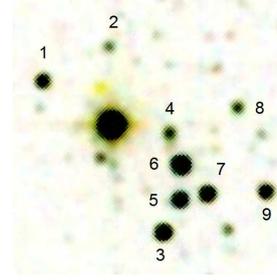}
\caption{Stars in the close angular vicinity ($1.2\arcmin \times 1.2\arcmin$; north is up, east to the left) of \hd\ (the brightest object in the snapshot) in $K_\mathrm{S}$ from 2MASS.}
\label{okoli}
\end{center}
\end{figure}

\begin{table}\scriptsize
\begin{center}
\tiny
\caption{Characteristics of stars in the direct vicinity of our target (sorted by $\alpha$), as shown in Figure \ref{okoli}. The columns denote, respectively, the 2MASS identifier, right ascension (J2000), declination (J2000), the $J$ band magnitude \citep{2mass}, and the magnitude difference to \hd, $\Delta$m. For the calculation of the latter value, we have adopted $J$\,=\,8.49\,mag for \hd, $C_2$ is the amplitude derived by the Eq.\,(\ref{true ampl}).}
\label{stars_surr}
\begin{tabular}{lc|ccccc}
  \hline
N&2MASS-ID & $\alpha$ (2000) & $\delta$ (2000) & $J$ & $\Delta$m & $C_2$\\
& & & & [mag] & [mag]& [mag] \\
 \hline
1&18511015-2423010	& 18 51 10.1	& --24 23 01	& 13.21	& 4.72  &0.7\\
2&18510866-2422503	& 18 51 08.7	& --24 22 50	& 14.77	& 6.28 & --\\
3&18510744-2423479	& 18 51 07.4	& --24 23 48	& 12.28	& 3.79 & 0.3\\
4&18510730-2423171	& 18 51 07.3	& --24 23 17	& 14.05	& 5.56 & 1.8\\
5&18510707-2423377	& 18 51 07.1	& --24 23 38	& 12.83	& 4.34 & 0.5\\
6&18510704-2423268	& 18 51 07.1	& --24 23 27	& 11.50	& 3.01 & 0.15\\
7&18510645-2423359	& 18 51 06.5	& --24 23 36	& 12.39	& 3.91 & 0.35\\
8&18510577-2423092	& 18 51 05.8	& --24 23 09	& 14.03	& 5.55 & 1.8\\
9&18510512-2423352	& 18 51 05.1	& --24 23 35	& 12.87	& 4.39 & 0.5\\
\hline
\end{tabular}
\end{center}
\end{table}

Before proceeding any further, we need to investigate the possibility that the observed peculiar light variability of \hd\ is caused by the presence of a variable star in the close angular vicinity that influenced the sky survey measurements.

The K2 aperture is $15\times23$ pixels, which corresponds to a field of $59\arcsec \times 92\arcsec$ on the sky. As has been pointed out in Sect.\,\ref{kepler_k2}, no significant contaminating sources were identified by an investigation of the corresponding plots from \citet{luger16}.

We used ASAS-SN observations for the inspection of nine objects in the close vicinity of \hd, listed in Table \ref{stars_surr} and indicated in Fig.\,\ref{okoli}, whose light variations could influence the results of the K2 photometry. The total contribution of these stars to the brightness measurements of \hd\ is non-negligible and amounts to 0.18\,mag. If any of the nearby stars was a periodic variable of sufficient amplitude $C_2$, as calculated by Eq.\,\ref{true ampl} (see the last column of Table\,\ref{stars_surr}), it could be the source of the observed secondary variability in the light curve of our target star.

ASAS-SN light curves of these objects usually span about two years, with up to 500 data points each. None of the investigated objects shows any sign of variability with upper limits between 0.01 and 0.05\,mag. Therefore, even if these stars show variability with amplitudes below the derived limits, no appreciable effect on the photometry of our target star is to be expected.

To verify and further investigate this, own CCD inspections (Sect. \ref{CCD}) were employed to search for traces of a faint star very close to our target. To this end, we fitted the point-spread-function (PSF) to all stars in the field (Fig.\,\ref{okoli}). Then we folded the individual PSFs with the model (i.e., the fitted PSF) of \hd\ for different distances from the center. As final step, we compared the derived synthetic PSFs with the observed one. From this procedure, we conclude that no star with a magnitude difference smaller than 5\,mag is located at a distance larger than 0.1\arcsec\ from our target star. Assuming the occurrence of 2.5 stars of the required parameters per square arcminute, we derive a probability of $3\times10^{-5}$ for such an arrangement. While the probability is small, it cannot be overlooked.

In summary, we conclude that the observable stars in the close angular vicinity do not contribute in any significant way to the observed light variations of \hd. Thus, if the observed light variability is due to an unresolved close companion, this object cannot be more distant than 0.1\arcsec\ from our target star and is not resolvable with the available instrumentation.

\subsubsection{Phenomenological model of the light curve}\label{phenmodel} %done

\begin{figure*}
\begin{center}
\includegraphics[width=0.34\textwidth]{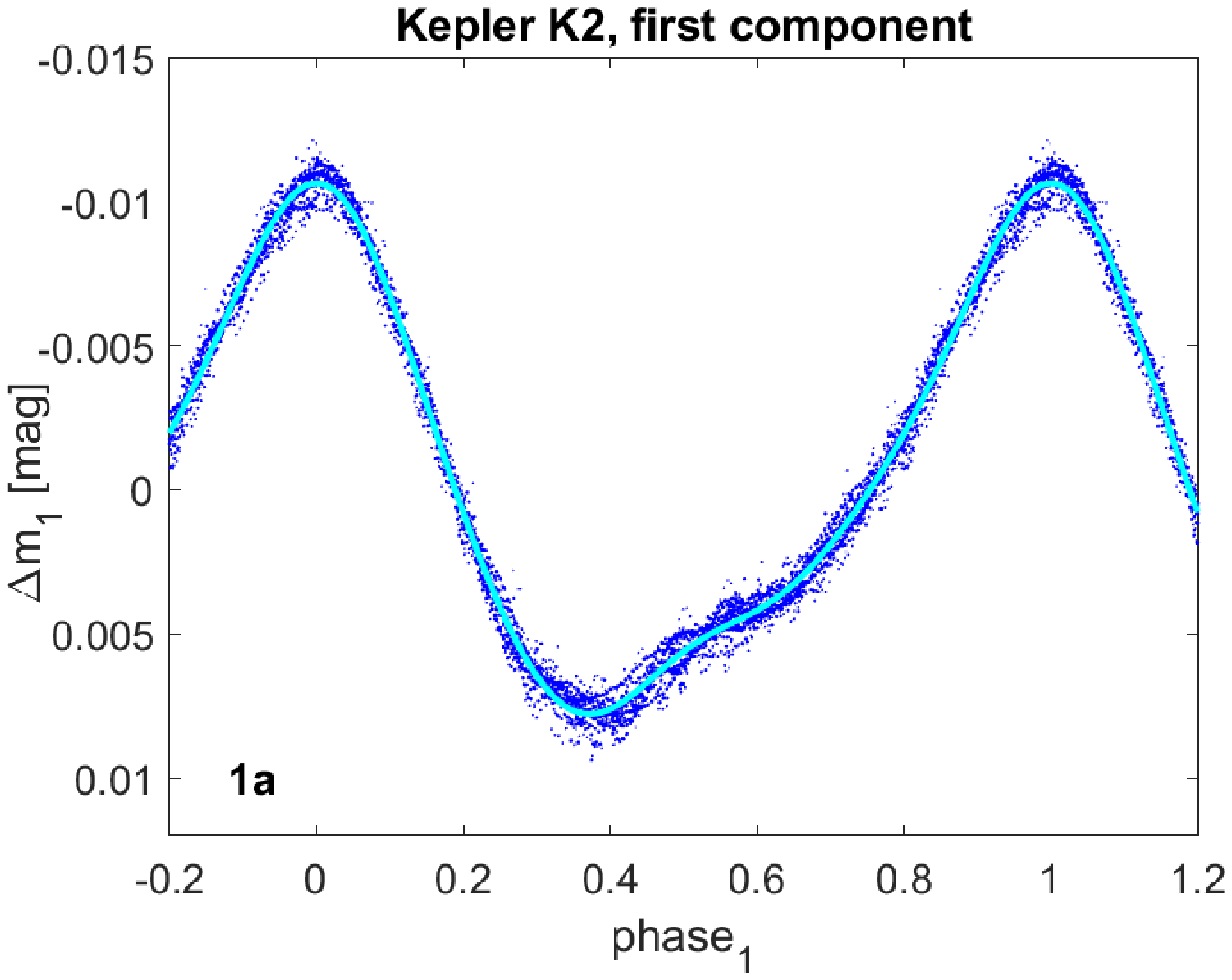}
\includegraphics[width=0.34\textwidth]{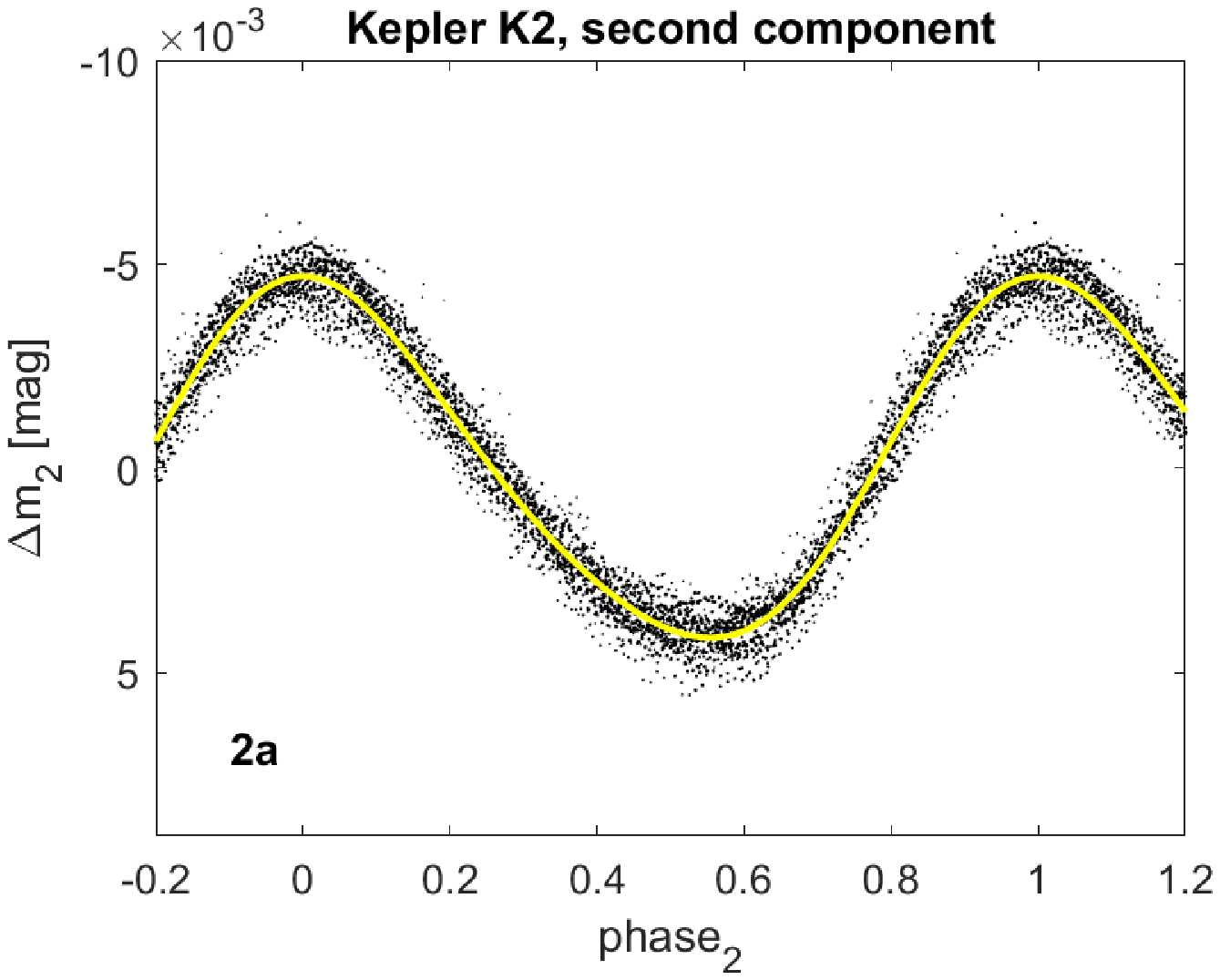}
\includegraphics[width=0.34\textwidth]{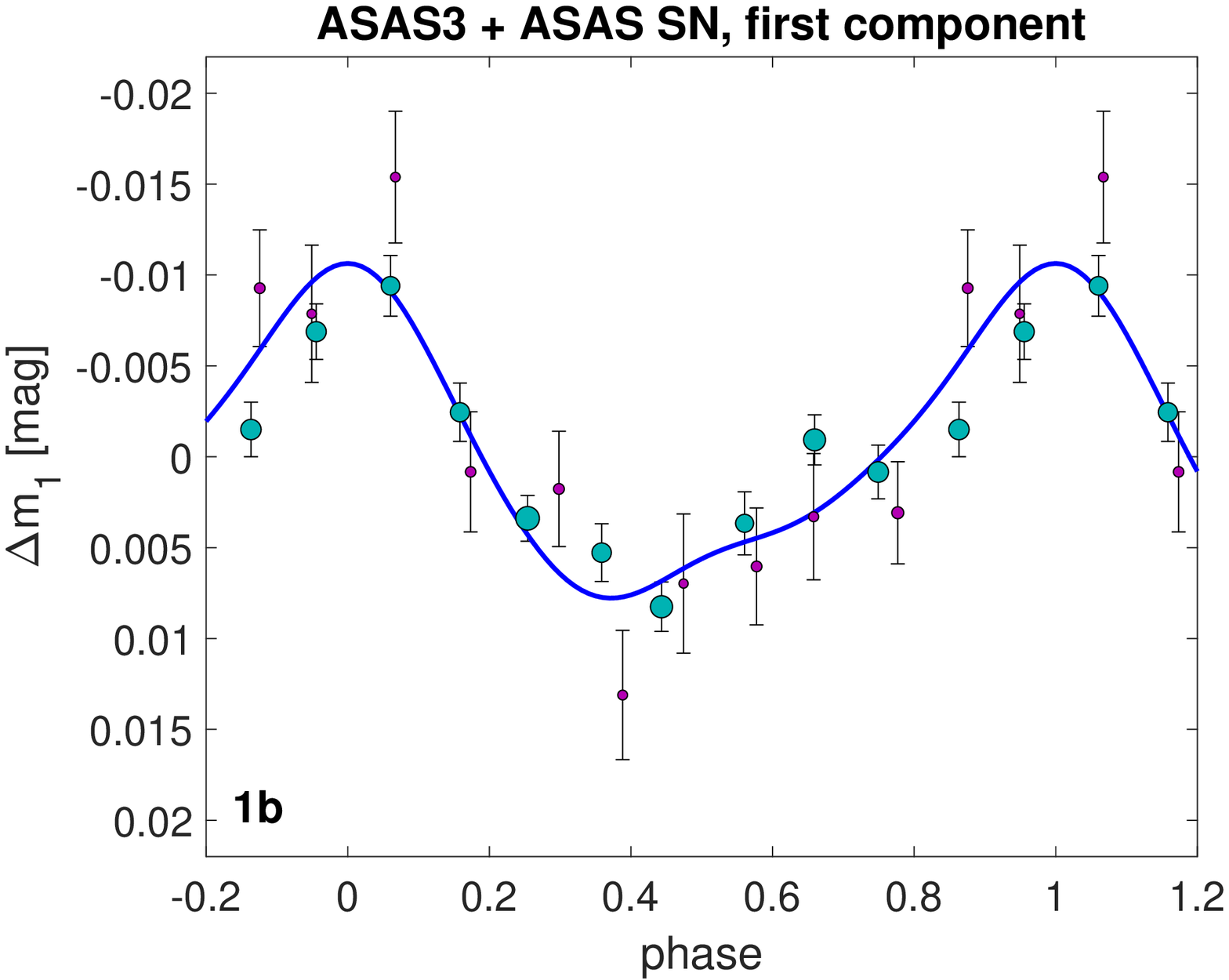}
\includegraphics[width=0.34\textwidth]{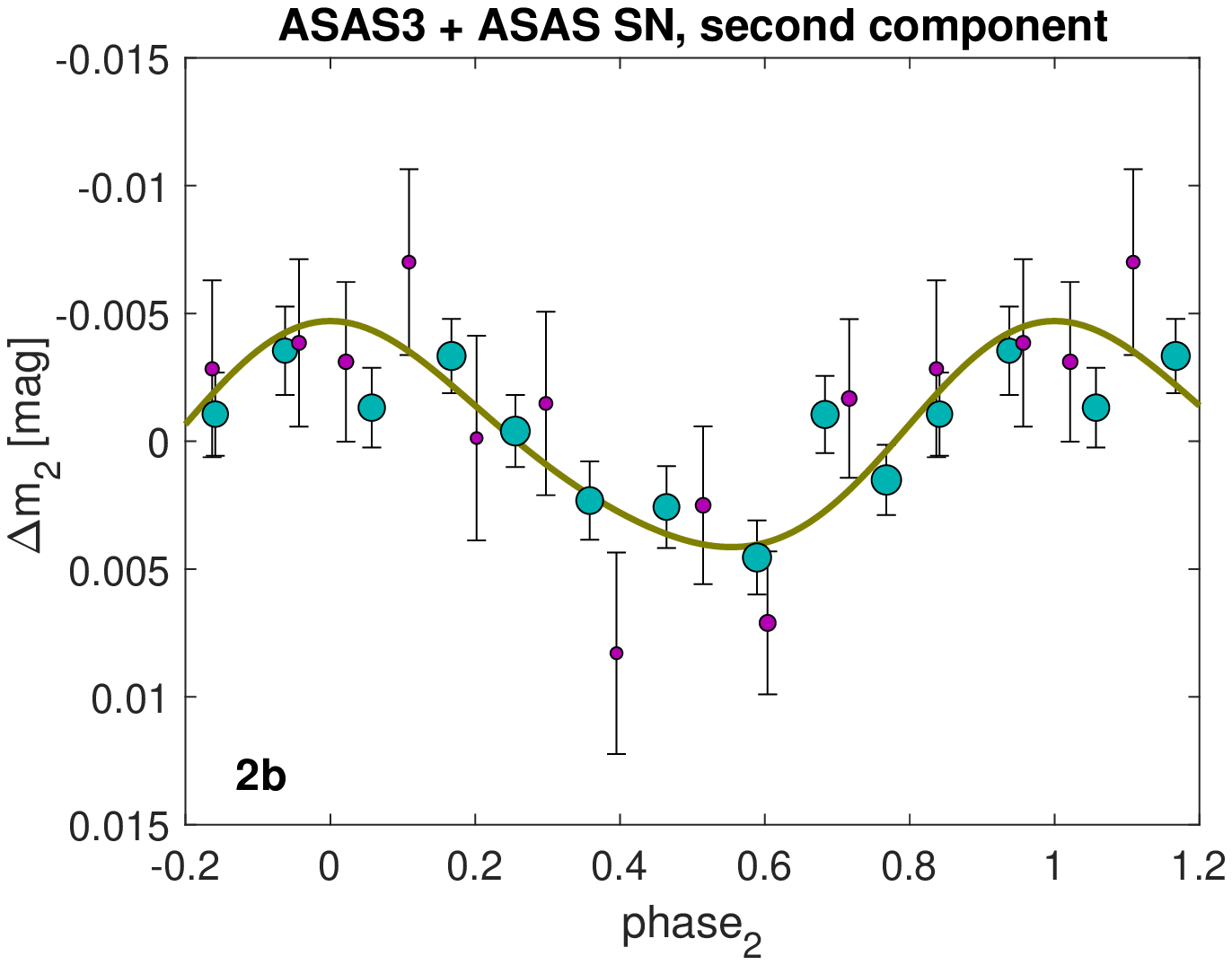}
\caption{Phased light curves plotted according to the linear ephemerides listed in Table\,\ref{tab} for the first light curve component (panels 1a and 1b) and the second light curve component (panels 2a and 2b). The phase diagrams in the upper panels (1a and 2a) were constructed using {\it Kepler}\,K2 data, while the diagrams in the lower panels (1b and 2b) were based on ASAS-3 and ASAS-SN data (green and lilac circles). The areas of the particular points correspond to their weights. Solid lines are the model fits according to Eq.\,({\ref{model}}).}
\label{kompot}
\end{center}
\end{figure*}

All photometric observations of \hd\ can be modeled by the linear combination of two periodic functions of the {\it phase functions} \citep[sum of epoch and phase, see e.\,g.][]{mik15} $\vartheta_1$ and $\vartheta_2$. The phase functions are defined using their particular periods $P_1,\ P_2$ and the times of basic maxima $M_{01},\ M_{02}$ with the relations in Eq.\,(\ref{fazovky}). The elementary light curves $F_1(\vartheta_1)$ and $F_2(\vartheta_2)$ are altogether single wave harmonic polynomials of different complexity with maximum at phases $\varphi_{1,2}=0$ \citep[see Eq.\,(\ref{model}) and][]{mik16}. The measure of variability of the individual components is the effective amplitude $B_{1,2}$, defined in Eq.\,(\ref{aeff}). The preprocessed K2 data are influenced by a long-term trend which is obviously an artefact of the K2 measuring and basic data treatment. This long-term trend can be expressed by a polynomial of the sixth order $D_{\rm K2}(t)$ of the time $t$.
\begin{eqnarray}
 &\vartheta_1=(t-M_{01})/P_1,\quad \vartheta_2=(t-M_{02})/P_2, \label{fazovky}\\
 &F_1=\sum_{j=1}^4\!b_{1\!j}\cos(2\pi j \vartheta_1)\!+\!\frac{b_{15}}{\sqrt{5}} \hzav{2 \sin(2 \pi \vartheta_1)\!-\!\sin(4 \pi \vartheta_1)}\label{model}\\
 & \quad +\,\frac{b_{16}}{\sqrt{70}}\hzav{3\sin(2\pi\vartheta_1)+6\sin(4\pi\vartheta_1)- 5\sin(6\pi\vartheta_1)},\nonumber\\
 & F_2=\sum_{j=1}^2 b_{2j}\cos(2\pi j \vartheta_2)+\frac{b_{23}}{\sqrt{5}}\hzav{2 \sin(2 \pi \vartheta_2)\!-\!\sin(4 \pi \vartheta_2)},\nonumber\\
 &F(t)=F_1(\vartheta_1)+F_2(\vartheta_2)+m_{0 i}+D_{\rm K2}(t_i),\nonumber\\
 & B_{1}= 2\sqrt{\sum_{j=1}^{6}\,b_{1j}^2},\quad B_{2}= 2\sqrt{\sum_{j=1}^{3}\,b_{2j}^2}\label{aeff},
\end{eqnarray}
where $b_{ij}$ are the parameters necessary for an adequate description of the light curves of both components, $m_{0i}$ is the mean magnitude of the data set to which the $i-$th observation belongs. $B_{1}$ and $B_{2}$ are the effective amplitudes of the first and second components according to the implementation of \citet{mik07}.

\begin{table}
\caption{Characteristics of the model functions $F_1$ and $F_2$. The effective amplitude $B_{1,2}$ is defined in Eq.\,(\ref{aeff}).}
\label{tab}
\centering
\begin{tabular}{l|cc}
  \hline
Parameter&  1-st component & 2-nd component  \\
 \hline
$ M_0$&2\,457\,329.2432(16)&2\,457\,337.3481(16)\\
$P  [d]$& 4.043\,61(5)& 2.111\,65(3)\\
$\dot{P}$& $2.4(6)\times10^{-7}$ & $4(4)\times10^{-8}$\\
$B_{1,2}$  [mmag]& 17.14(3)& 8.77(4)\\
\hline
\end{tabular}
\end{table}

All 25 free parameters of our photometric variability model of \hd\ and the corresponding uncertainties were determined by robust regression (RR) as implemented in \citet{mik03,mik11}, which exploits the well-established procedures of the standard weighted least squares method and eliminates the influence of outliers. Instead of the usual $\chi^2$, there is minimised the modified quantity $\chi^2_{\rm r}$, defined as follows:
\begin{eqnarray}
& \chi^2_{\rm r}=\sum_{i=1}^n\,\zav{\frac{\Delta y_i} {\sigma_{\mathrm{r}i}}}^2; \quad \mathrm{where}\ \ \sigma_{\mathrm{r}i}=\sigma_i\ \exp\hzav{\frac{1}{2}\zav{\frac{\Delta y_i}{4\,\sigma_i}}^4},\\
&\chi^2_{\mu}=1.06\ \frac{\chi^2_{\mathrm{r}}}{n_{\rm r}-g};\quad \mathrm{where}\ \
n_{\rm r}=1.02\ \frac{\sum\,\sigma_{\mathrm{r}i}^{-2}}{\sum\,{\sigma_{i}}^{-2}}, \label{chimi}
\end{eqnarray}
where $\Delta y_i$ is the difference between the observed $i$-th measurement $y_i$ and the model prediction $f(t_i,\boldsymbol{\gamma})$, which is the function of the time of the measurement $t_i$ and the vector of the free model parameter $\boldsymbol{\gamma}$. $\sigma_i$ is the estimate of the uncertainty of determination of the $i$-th measurement, while $\sigma_{\mathrm{r}i}$ is a RR modified value of $\sigma_i$. The estimate of the common relative $\chi^2_{\mu}$ and the number of measurements without outliers $n_{\rm r}$ are given by Eq. (\ref{chimi}).

The uncertainties of the derived parameters $\boldsymbol{\gamma}$ were determined by standard LSM regression techniques. The most interesting parameters and their functions are listed in Table\,\ref{tab}.

The model fits the observed light curve with a weighted uncertainty of only 0.5\,mmag (see Fig.\,\ref{detrlc}). It is also able to explain the observed frequency spectra shown in Figs.\,\ref{periodog} and \ref{periodog2}. The phase plots of the individual light curve components are also well compatible with the models (see Fig.\,\ref{kompot}). Their shape will be discussed in detail below (Sect.\,\ref{LCs_interpret}). We also tested the long-term stability \citep{mik16} of the periods $P_1$ and $P_2$ and find $\dot{P}_1=2.4(6)\times10^{-7}$ and $\dot{P}_2=4(4)\times10^{-8}$, indicating that long-term changes are either absent or insignificant.

\subsubsection{Interpretation of the light curve components} \label{LCs_interpret} %done

The primary light curve component ($P_1=4.043\,61(5)$\,d) has a single-wave, slightly humped shape that is characteristic of the rotational light changes of mCP stars with two or more persistent bright photometric spots \citep[][and references therein]{jagelka15}. The areas of different contrast partly overlap and are asymmetrically located with photo-centers at phases $\varphi_{011}=0.02$ and $\varphi_{012}=0.85$.

The secondary light curve component ($P_2=2.111\,65(3)$\,d) also has a single-wave shape, which shows a steeper ascending branch (see Fig.\,\ref{kompot}). This type of light curve is characteristic for specific types of pulsational variables and also for mCP stars with asymmetrically located photometric spots.

This leaves us with two groups of possible scenarios explaining the light curve of \hd:
\begin{enumerate}
\item a single star showing ACV variability and single-mode classical pulsation or, perhaps, differential latitudinal rotation,
\item an optical or physical binary with different combinations of variability.
\end{enumerate}

The following analysis of the spectral changes and radial velocity variations provides important constraints on these scenarios, which are discussed in detail in Sect.\,\ref{discus}.

\begin{figure*}
\begin{center}
\includegraphics[width=0.40\textwidth]{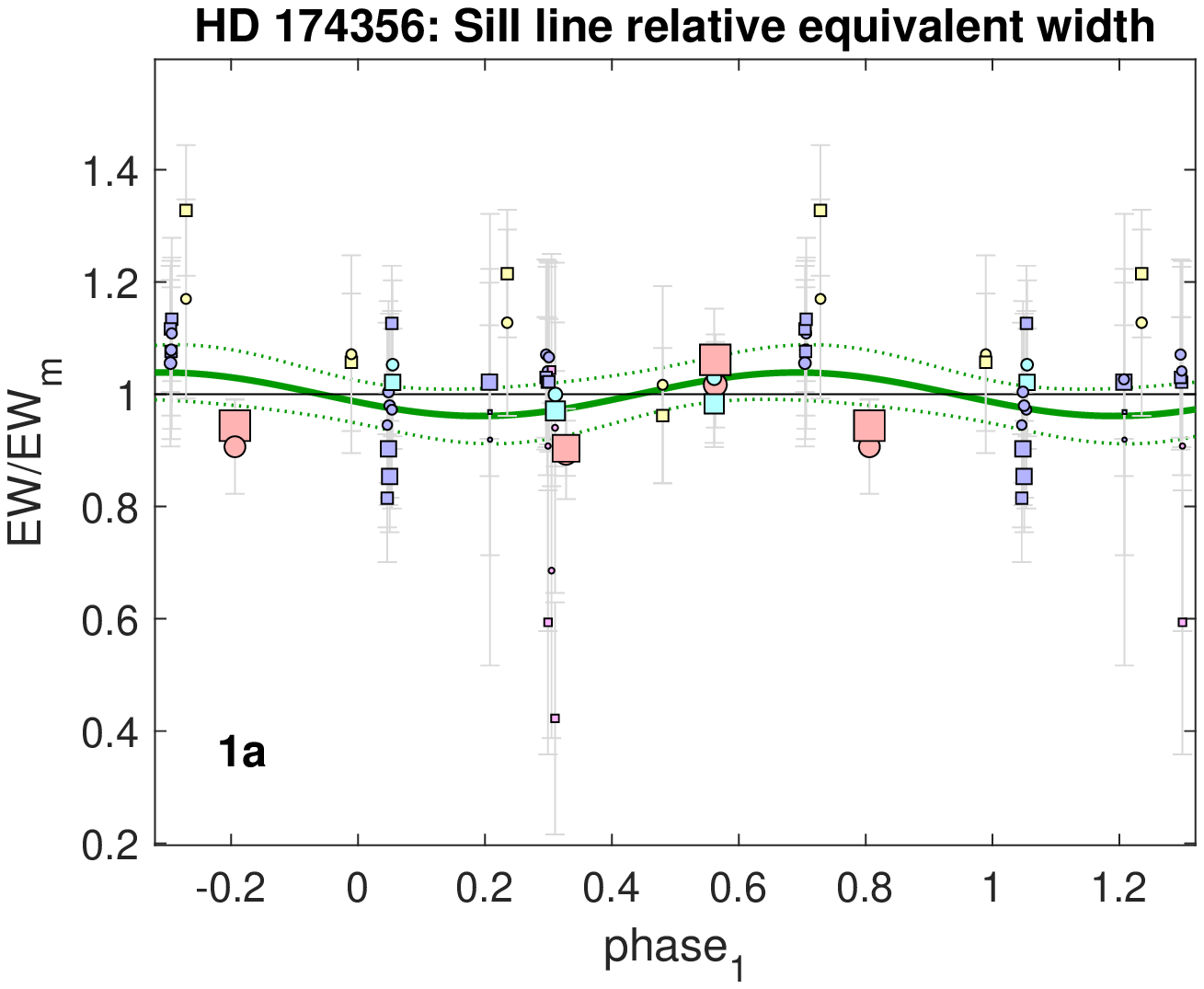}
\includegraphics[width=0.40\textwidth]{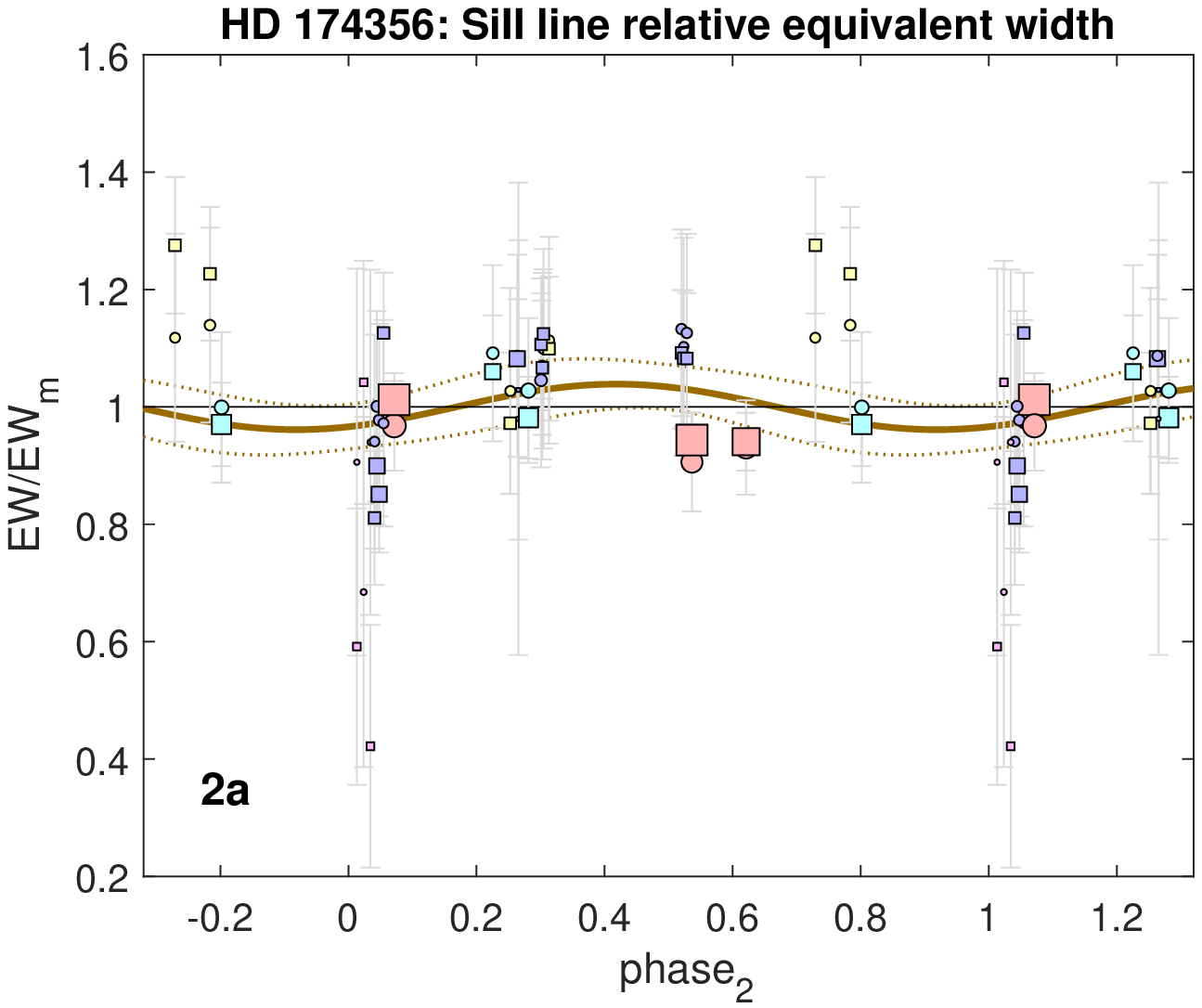}
\includegraphics[width=0.40\textwidth]{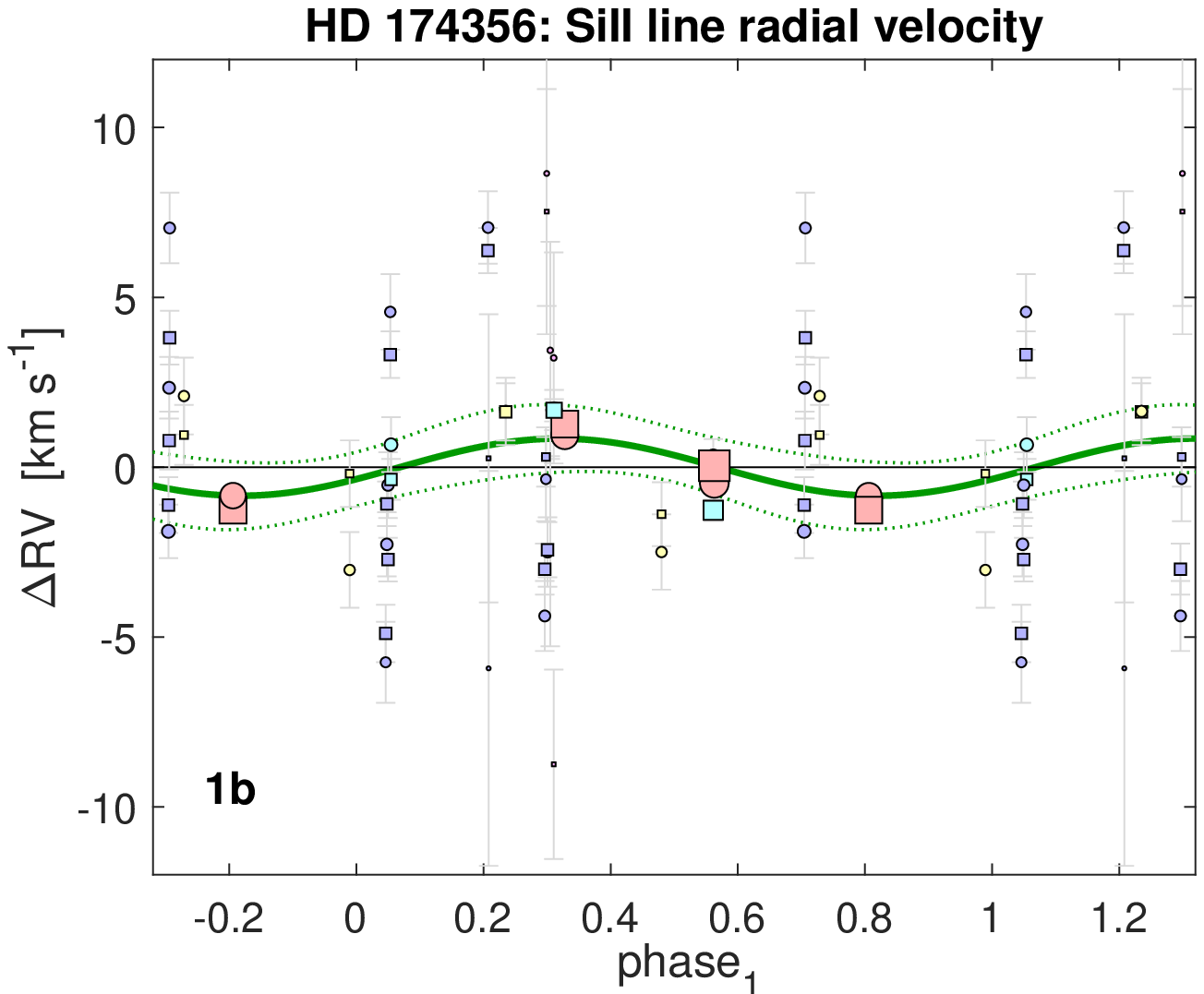}
\includegraphics[width=0.40\textwidth]{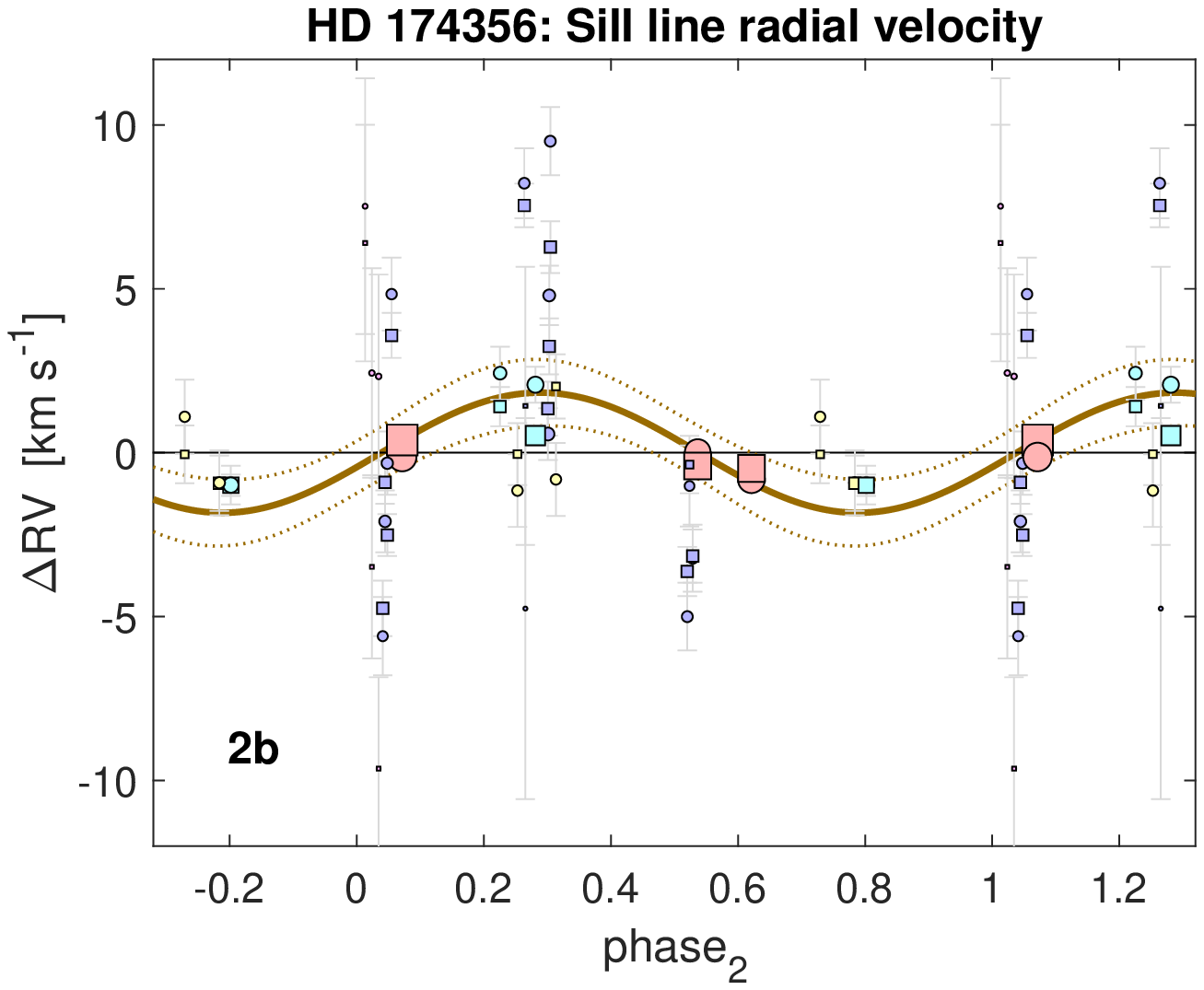}
\includegraphics[width=0.40\textwidth]{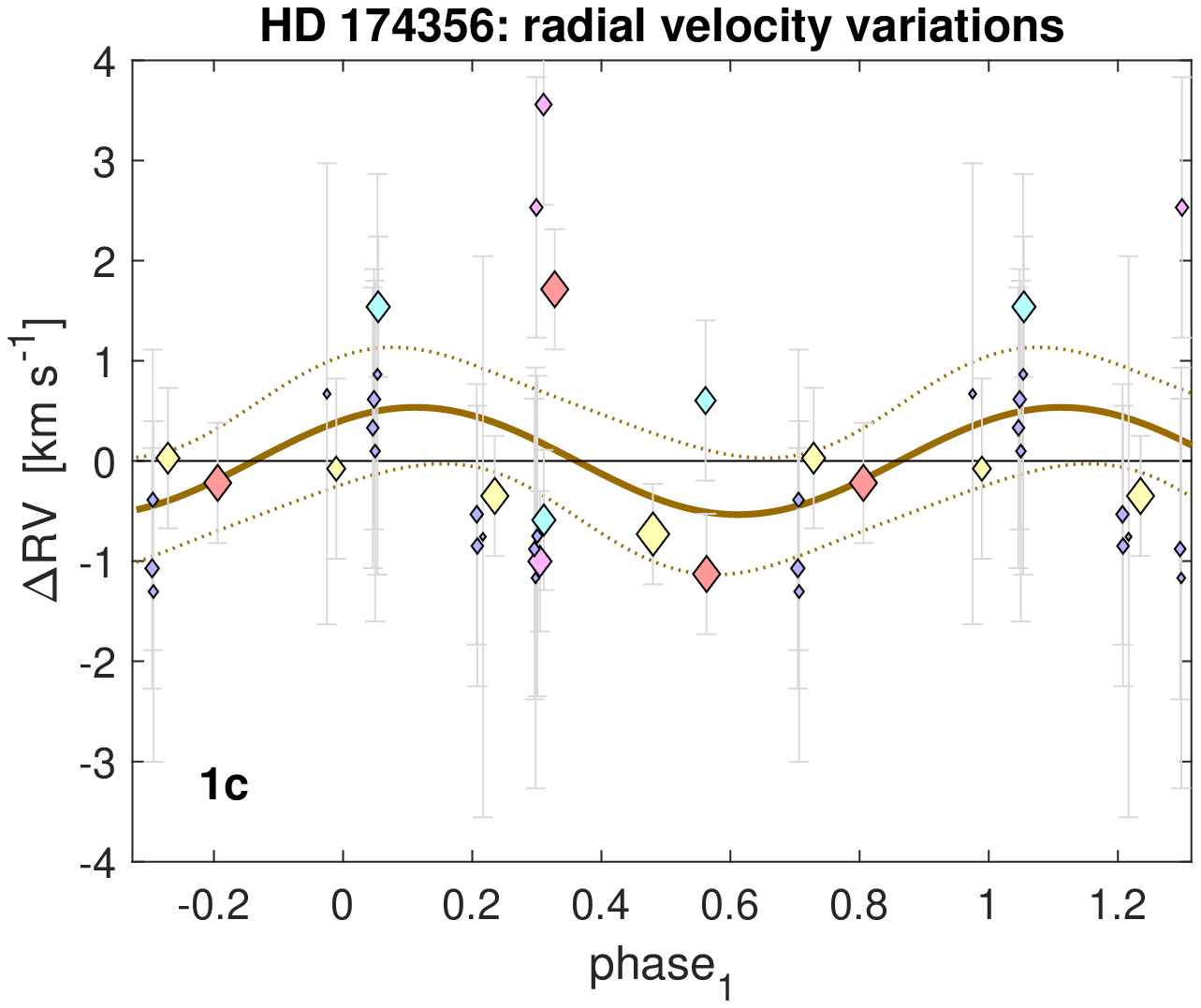}
\includegraphics[width=0.40\textwidth]{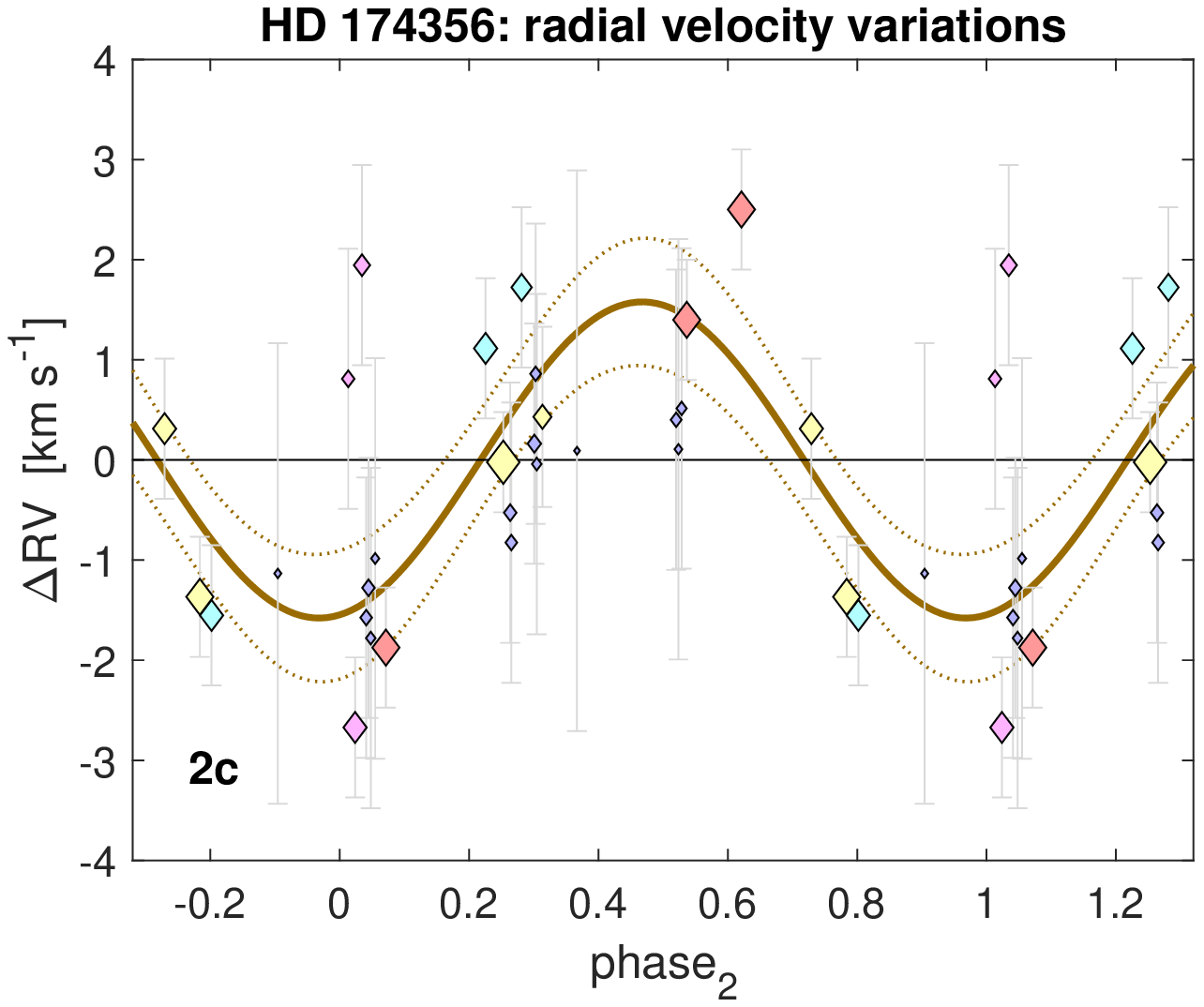}
\caption{Differences of the mean relative equivalent widths $EW_{\rm {rel}}$ (upper panels 1a+2a) and radial velocities $RV$ (middle panels 1b+2b) of two \ion{Si}{II} lines plotted versus the phases of the first (left panels) and the second (right panels) light curve components. Bottom panels (1c+2c) show the differences of radial velocity in relation to the mean value of $RV = +12.5(6)$\kms. In all panels, the areas of markers correspond to their weights and the green/brown lines indicate the fit of the phenomenological models according to (from top to bottom panels) Eqs. \ref{EWhar}, \ref{RVhar} and \ref{tothar}, corrected for the variations in the other periodicity. Dotted lines indicate the modeled 1 $\sigma$ deviation from the mean harmonic course versus the phase of the first light curve component. Sources are colour-coded: magenta -- PST2, red -- HARPS, yellow -- HIDES, cyan -- FERROS, blue -- CASLEO + SOAR (cf. Table\,\ref{SiEWRV}).}
\label{EWSi}
\end{center}
\end{figure*}

\subsection{Spectral variability} \label{rvs} %done

In the following, we analyse the variations of the radial velocities and equivalent widths of selected spectral lines. In agreement with the literature results, we identify the main photometric period $P_1$ as the rotational period of \hd. If the light variability is due to the presence of abundance spots, which we assume for CP2 stars, we expect to find variability in the equivalent widths and radial velocities of the lines of the corresponding overabundant elements. Silicon, which is significantly overabundant in our target star (see Sect. \ref{atmo}), tends to play a major role in the light variability of late-type Bp stars. We therefore concentrated on the radial velocities ($RV$) and equivalent widths ($EW$) of two selected \ion{Si}{II} lines ($\lambda$5041\,\AA\ and $\lambda$5056\,\AA), which were measured with the program SPLAT-VO \citep{Skoda2014}. The results of the measurements are listed in Table\,\ref{SiEWRV}, which also includes photometric phases calculated with the parameters listed in Table\,\ref{tab}.

The $EW$ and $RV$ changes were searched using the simplest possible model admitting simultaneous variability in both detected photometric periods (Eqs. \ref{EWhar} and \ref{RVhar}). A similar model (Eq. \ref{tothar}) was also applied to the differential radial velocity variations $\Delta RV$. All 27 spectra obtained in 2017 and 2019 (Sect. \ref{Spectroscopy}) were used for the analysis.

\begin{eqnarray}
& EW(t)  = \overline{EW}\,\hzav{1 + \sum_{j=1}^2 h_{j1} \cos(2\pi\vartheta_j)+h_{j2}\sin(2\pi\vartheta_j)},\label{EWhar}\\
&  RV(t)=\overline{RV}+\sum_{j=1}^2 g_{j1} \cos(2\pi\vartheta_j)+g_{j2}\sin(2\pi\vartheta_j),\label{RVhar}\\
&  \Delta RV(t)=\overline{\Delta RV}+\sum_{j=1}^2 a_{j1} \cos(2\pi\vartheta_j)+a_{j2}\sin(2\pi\vartheta_j),\label{tothar}
\end{eqnarray}
where $\vartheta_1$ and $\vartheta_2$ are the phase functions calculated according to Eq. (\ref{fazovky}). The parameters of the models and their uncertainties, as determined by bootstrapping, are listed in Table \ref{spepar}.

The quality of the spectra used for the analysis is very different and strong scattering around the modeled interdependencies is evident in Figure\,\ref{EWSi}. We therefore caution that the modeling results, although obtained by bootstrapping, have to viewed with caution.

Considering the uncertainties, no significant equivalent width changes in the measured silicon lines can be demonstrated (Fig.\,\ref{EWSi}, upper panels 1a+2a). There is, however, some indication of corresponding radial velocity variations (middle panels 1b+2b). Compared to the uncertainties, the observed amplitudes are small (1.7(1.3)\,\kms\ and 3.7(1.6)\,\kms\ as phased with, respectively, the periods of the first and second light curve components). Furthermore, reliable data is distributed very unevenly across the phase curve shown in panel 2b. However, the inflection point of the $\Delta RV$ curve plotted against $\varphi_1$\ (panel 1b) occurs at phase $\varphi_1$\,=\,$0.07(0.12)$, which roughly suits the expectations and agrees with the assumption of an uneven atmospheric distribution of silicon. We caution, however, that the observed changes in the measured silicon lines are small.

More information can be gleaned from the radial velocity curve derived from the measurements of the shift of the spectrum relative to the mean spectrum of the $\Delta RV$ values (Fig. \ref{EWSi}, bottom panels 1c+2c). According to phase $\varphi_1$, the point of inflection occurs at $\varphi_1 = 0.86 (11)$, which agrees very well with the position of the light curve maximum as approximated by a simple sine wave. These variations are in agreement with the assumption of a large spot of metals with lines dominating in the optical spectral region that is responsible for the observed light changes.

Furthermore, a relatively large amplitude of 3.1(1.1)\,\kms, which is very likely real (certainty of 98\,\% according to the shuffle test; \citealt{mikper15}), is detected in the $\Delta RV$ measurements with respect to period $P_2$. A minimum is observed at phase $\varphi_2 = -0.03(4)$, at the moment of light maximum of the secondary component.

We stress that the simple explanation that the observed RV changes reflect the orbital motion of a gravitationally bound primary with another star that is invisible in the spectrum does not give an astrophysically acceptable solution, as the proposed secondary star would have to orbit below the surface of the primary component. This assumption is therefore invalid. Similarly, these results are also not in agreement with the hypothesis that \hd\ is an optical binary, i.e. a chance alignment of two physically unbound stars of different distances, since it is impossible for a remote optical companion to influence the spectrum of the other star.

A plausible solution would be obtained, however, by assuming that the radial velocity changes with period $P_2$ are the results of pulsation in a single star, because for single-mode pulsators, maximum brightness occurs at a phase close to the phase of minimal radial velocity. This hypothesis, among others, is further discussed below.

\section{Discussion}\label{discus}   

\hd\ appears to be a standard single CP2 star, with the exception of its peculiar light curve that can be decomposed into two independent components. The first component is in agreement with rotational modulation caused by the presence of abundance inhomogeneities on the surface of an mCP star, which is partly supported by the spectral characteristics of our target star (see Sect.\,\ref{rvs}).

When considering the secondary light curve component, it is important to bear in mind that the search for the spectral lines of a possible secondary stellar component was entirely unsuccessful and the available evidence points to \hd\ being a single CP2 star (cf. Sect.\,\ref{rvs}). Any object contributing to the photometry of \hd\ must therefore be a physically unbound, line-of-sight companion situated at a distance of 0.1\arcsec\ or less from our target star (cf. Sect.\,\ref{vicinity}). However, the well-defined radial velocity curve of \hd\ folded on period $P_2$ (see Fig.\,\ref{EWSi}, panel 2c) is a strong argument against the object being an unresolved pair of periodically variable stars. Nevertheless, for the sake of completeness, binarity hypotheses are also included in the following discussions.

\subsection{Double star hypotheses}\label{double}

\subsubsection{Unresolved pair of stars} \label{unresolvedb}

Let us first assume that \hd\ is an unresolved physical or optical double star. Unfortunately, for an investigation into the nature of the individual binary components, we can only rely on the shape of the light curves (see Fig.\,\ref{kompot}). As has been pointed out, additional information on the \textit{true} effective amplitudes of the individual light curve components $C_1$ and $C_2$ cannot be derived from the \textit{observed} effective amplitudes $B_1$ and $B_2$ (Table\,\ref{tab}) without knowledge of the magnitude difference $\Delta m$ of their mean apparent (observed) magnitudes $\overline{m}_1$ and $\overline{m}_2$.

\begin{figure}
\begin{center}
\includegraphics[width=0.38\textwidth]{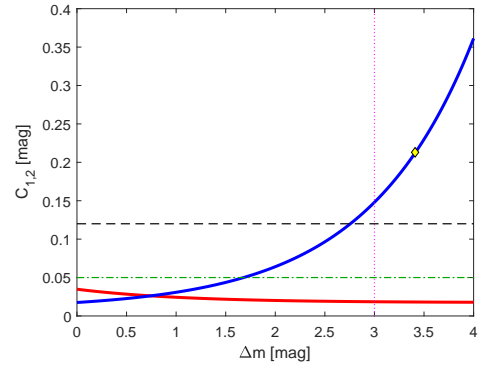}
\caption{Relation between the true effective amplitudes of both components, $C_{1,2}$, and the difference of their mean magnitudes $\Delta m$ (red line - the first component, blue line - the second one). Coloured lines indicate: magenta dotted line -- limit of visibility/invisibility of the spectral lines of the secondary component; dashed black line -- the maximum observed amplitude of CP2 stars $(C_2\sim 0.12$\,mag); dashed-dotted green line -- the typical amplitude of CP2 stars. The diamond indicates the position of the first-overtone Cepheid V397 Car (see Sect.\,\ref{cepheids}.)}
\label{zavist}
\end{center}
\end{figure}

In the case of physical double stars, the difference between the apparent magnitudes $\Delta m=\Delta M=M_2-M_1$ also expresses the difference between the absolute magnitudes. However, for optical pairs whose components might suffer different amounts of interstellar extinction, we need to take into account the uneven distance moduli $\mu_1$ and $\mu_2$ and the different extinction values $\Delta A_V= A_{V}(\mu_2) - A_V(\mu_1)$. We then obtain the more general relation:
\begin{equation}\label{extinkce}
\Delta m=\Delta M + \mu_2-\mu_1 + \Delta A_V(\mu_1,\mu_2).
\end{equation}

After some algebra, we derive the following versatile relations:
\begin{align}
&\Delta m=\overline{m}_2-\overline{m}_1; \quad \overline{m}_{1,2}=\overline{m}+2.5\,\log_{10}\zav{1+10^{\mp0.4\,\Delta m}};\\
&\displaystyle B_2=2.5\,\log_{10}\zav{\frac{10^{0.4\,\Delta m}+10^{0.2\,C_2}}{10^{0.4\,\Delta m}+10^{-0.2\,C_2}}};\label{secampl}\\
&\displaystyle \Delta m=2.5\,\log_{10}\hzav{\frac{10^{0.2\,C_2} -10^{(0.4\,B_2-0.2\,C_2)}}{10^{0.4\,B_2}-1}};\nonumber\\
&\displaystyle C_2\doteq\ 2.5\!\log_{10}\!\hzav{\frac{10^{0.4(B_2\!+\!\Delta m)}\!+\!2\cdot\!10^{0.4\,B_2}\!-\!10^{0.4\,\Delta m}}{2-10^{0.4(B_2+\Delta m)}+10^{0.4\,\Delta m}}}. \label{true ampl}
\end{align}
After interchanging the indices ($1\!\leftrightarrows\!2$), the same relations are valid also for the variations of the first component. The basic relation between the true effective amplitude of the secondary star $C_2$ and the difference of the mean magnitudes of the components $\Delta m$ for the observed effective amplitude of the secondary component, $B_2=8.77$\,mmag, is plotted in Figure\,\ref{zavist}.

Since we know the period (Table \ref{tab}) and shape (Fig.\,\ref{kompot}) of the secondary light curve component, we can narrow down the origin of the variability. Using the relations derived in Eq.\,(\ref{secampl}) also enables us to discuss the nature of the secondary component quantitatively. We also know that the characteristics of the light curve have remained stable over more than 30 cycles, and the period has been constant for about 17 years. Therefore, variable stars prone to exhibiting significant period jitter or light curve changes have not been considered.

\subsubsection{Unresolved mCP star companion}

Since both light curve components of \hd\ are in agreement with mCP star light curves (see Sect.\,\ref{LCs_interpret}), we will focus first on discussing the scenario of an unresolved optical pair of mCP stars. Using the equations provided in Sect.\,\ref{unresolvedb}, we can estimate the true effective amplitudes $C_1(\Delta m)$ and $C_2(\Delta m)$ as functions of the difference of the observed mean magnitudes $(\Delta m=m_2-m_1)$ (see the red and blue lines in Fig.\,\ref{zavist}).

Unfortunately, the parameter $\Delta m$ cannot be directly derived from photometry alone. However, the fact that no spectral lines of the secondary component are visible in the spectrum of \hd\ implies that a hypothetical secondary component should be fainter by three or more magnitudes $(\Delta m \geq 3$\,mag; cf. Fig. \,\ref{zavist}). To modulate the light variations of the primary star (effective amplitude $B_1=17.14$\,mmag) by the observed amplitude of $B_2 =8.7$\,mmag, the amplitude of the light changes of the secondary star should, according to Eq.\,(\ref{secampl}), exceed 0.15\,mag (see Fig.\,\ref{zavist}).

However, rotational light changes of such a large amplitude in $V$ have never been observed in mCP stars \citep[][]{mikzoo07}. With an amplitude of 0.12\,mag in the $V$ or $y$ filters \citep[][and references therein]{dukes18}, HD 215441 is known as the mCP star with the largest photometric amplitude in the optical region. Even if the inclination of this star was not $i=67.5^{\circ}$ \citep{khokhlova97} but $i=90^{\circ}$, the amplitude would remain safely under the 0.15\,mag limit. While it is not impossible that mCP stars with amplitudes exceeding 0.12\,mag ($V$) exist, the assumption of such a record-breaking object as companion to our target star seems rather unlikely. Besides, mCP stars with large amplitudes generally also show a high degree of peculiarity in their spectra, which is not in agreement with the mildly peculiar spectrum of \hd.

While we cannot totally exclude this scenario, it seems very unlikely that the hypothetical companion to \hd\ is a classical mCP star.

\subsubsection{First-overtone Cepheid companion}\label{cepheids}

Considering the large amplitude of the secondary light curve component, we next investigated the possibility that the secondary is a short period pulsating variable. Among the regular pulsators, RR\,Lyrae stars can be ruled out on grounds of the observed period of 2.111\,65(3)\,d, which puts the star into the realm of the Cepheids \citep{catelan15}. Type II Cepheids consist of several subclasses, with the BL\,Herculis (also termed CWB) stars exhibiting periods shorter than 8 days \citep{gcvs}. However, type II Cepheids show a distinct light curve progression with period, and the light curves of BL Her stars with periods at around 2 days are characterised by obvious secondary bumps, which is not in agreement with the shape of the secondary light curve component. Furthermore, type II Cepheids are known to exhibit significant cycle-to-cycle variations in their light curves \citep{Schmidt2009}.

\begin{figure}
\begin{center}
\includegraphics[width=0.35\textwidth]{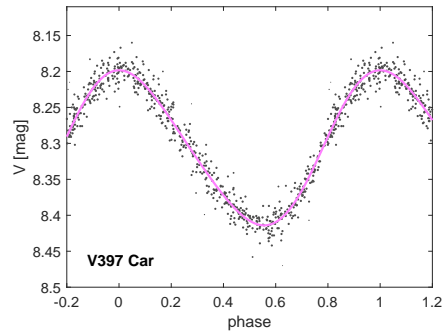}
\caption{Phase diagram of the first overtone short period Cepheid V397 Carinae, constructed using ASAS-3 data. Concerning period and light curve shape, V397 Car is a twin of the hypothetical secondary component of \hd.}
\label{Cepheid}
\end{center}
\end{figure}

Classical Cepheids, on the other hand, exhibit light and radial velocity variations that repeat very precisely. Their light curves show a characteristic relationship between the pulsation period and the presence and position of a bump (the so-called ``Hertzsprung Progression''; e.g. \citealt{bono00}). Interestingly, the observed period and shape of the secondary light curve component are in line with a Cepheid pulsating in the first overtone \citep{evans2015}. Figure \ref{Cepheid} shows the light curve of the short-period first overtone Cepheid V397 Carinae (TYC 8955-1015-1), whose similarity with the light curve of the secondary component of \hd\ is striking (see Fig.\,\ref{kompot}).

As it follows from \citet{Klagyivik2009}, the effective amplitudes $C_2$ of first overtone Cepheids near the short-period border are considerably smaller than those of other Cepheids and typically of the order of 0.15 -- 0.4 mag. Using Eq.\,(\ref{secampl}), we find that $\Delta m$ should be in the range from 3.02 to 4.13 mag. Following this scenario, it is possible to estimate the geometric distance $d$ of the objects by employing the period-luminosity correlation for first overtone Cepheids from \citet{Ripepi2012} and the \citet{green15} extinction maps in the interval $9<d_2<15$\,kpc. This leads to the derivation of the following distance moduli and distance intervals for the secondary component: $14.8<\mu_2<15.9$ mag and $9<d_2<15$\,kpc. \hd\ is located in the first Galactic quadrant with [$l$,$b$]\,=\,[10.93,$-$10.79], which means that the hypothetical first-overtone Cepheid candidates would be located behind the Galactic center and from 1.7 to 2.9 kpc below the Galactic plane. However, Cepheids are young objects of the disk population \citep{Mor2017} and hardly exceed distances of 0.5\,kpc above or below the Galactic disk \citep{Fernie1995}. Only run-away stars \citep{Kenyon2014} may attain greater distances, but these are very rare among Cepheids. Therefore, the assumption of a Cepheid companion, while being able to explain the observed secondary light curve component, is highly unlikely.

In summary, we conclude that the assumption of a binarity scenario to explain the peculiar light changes of our target star is highly improbable.

\subsection{Differential latitudinal rotation of an mCP star}

It is commonly assumed that stars of the upper main sequence are rigid body rotators. However, \citet{Reiners2004} were the first who suggested that a few A-type stars show clear signs of differential rotation. Using CoRoT data, \citet{Degroote2011} presented strong evidence for the existence of spots and differential rotation in the B8/9 star HD\,174648.

\citet{bowman18} suggested that the observed light variations of \hd\ might be the result of latitudinal surface differential rotation of a single mCP star. However, according to the current state-of-knowledge and as shown by the example of the carefully monitored star HD 37776 \citep{mik18}, there is no evidence that mCP stars show longitudinal differential rotation and we do not see reason to believe that \hd\ forms an exception. The proposed model would necessarily lead to the dissolution of the chemical surface structures, and thus to the gradual disappearance of the rotationally modulated variations. Besides, it fails to explain the strictly periodic nature of the secondary light curve component. We therefore feel safe in rejecting this hypothesis.

\subsection{\hd\ as \textbf{S}lowly \textbf{P}ulsating \textbf{B}-type star} \label{pulsation_SPB}

It is well known that Slowly Pulsating B-type (SPB) stars share the same location in the Hertzsprung-Russell diagram (HRD) as mCP stars \citep{2007A&A...466..269B}. \emph{They are high radial order} g-mode pulsators with periods of 0.5 to 5 days, which usually show rich and complicated frequency spectra. In the case of \hd, we have derived only a single mode, but this may be connected with the detection threshold and the presence of other low-amplitude frequencies cannot be excluded. This situation is reminiscent of the SPB star $\nu$ Eri, for which only two pulsation modes were detected in ground-based data \citep{2004MNRAS.347..454H} and spectroscopic \citep{2004MNRAS.347..463A} campaigns. However, high-accuracy space-based observations, \citep{2017MNRAS.464.2249H} revealed seven low amplitude g-modes in this star.

We here investigate the possibility that the secondary light curve component could be explained by the presence of g mode pulsation in our target star.

\subsubsection{Stellar parameters from the HRD}

The location of \hd\ within the HRD is shown in Fig.~\ref{fig:HR}. The error box was determined assuming effective temperature $\log{T_{\rm{eff}}}$\,=\,$4.1206(99)$. The adopted luminosity, $\log{L/\rm {L}}_{\sun}$\,=\,$2.834(83)$, was calculated using the Gaia parallax $\pi=1.18(11)$\,mas \citep{brown16,gaiacol18,2018A&A...616A...2L}. The bolometric correction, BC=$-0.924(57)$, was adopted from \citet{Flower96}.

Evolutionary tracks for masses from 4.10 to 4.70\,M$_{\sun}$ were calculated with the Modules for Experiments in Stellar Astrophysics (MESA) code \citep[][and references therein]{MESA4}. We also applied the MESA Isochrones and Stellar Tracks (MIST) configuration files \citep{MIST0,MIST1}. We assumed metallicity $Z=0.015$ and initial hydrogen abundance $X_{\rm{ini}}=0.7$. Furthermore, OPAL opacity tables \citep{OPAL} supplemented with the data provided by \cite{2005ApJ...623..585F} for the low temperature region and the solar chemical element mixture as determined by \citet{Asplund2009} were used for the computations. In Fig.~\ref{fig:HR}, the results derived for two different values of the exponential overshooting parameter from the hydrogen-burning convective core, $f_{\rm{ov}}=0.01$ and 0.02, are shown.

\begin{figure}
\begin{center}
	\includegraphics[width=0.49\textwidth]{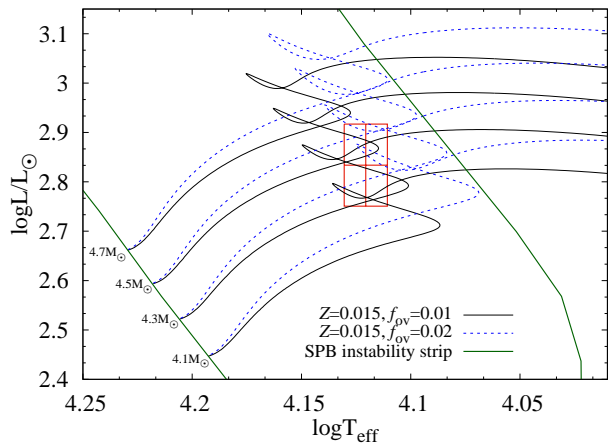}
\caption{HR diagram with stellar evolutionary tracks, the position of \hd, and its corresponding error box. Evolutionary models were calculated for masses from 4.1  to 4.7\,M$_{\sun}$ with a step-size of 0.20\,M$_{\sun}$. We assumed metallicity $Z=0.015$, initial hydrogen abundance $X_{\rm {ini}}=0.7$, and two values of the exponential overshooting parameter $f_{\rm {ov}}=0.01$ and 0.02.}
	\label{fig:HR}
\end{center}
\end{figure}

From the position of the star in the HRD, it is not possible to unambiguously derive its evolutionary stage. \hd\ can be a main-sequence star that is either in the Contraction Phase (CoP) or situated After the Loop (AtL). The expected evolutionary changes of the rotational frequency are smaller than the accuracy limit of the available measurements and therefore not suitable to solve this problem.

However, the HRD can be used for deriving the radius, $R$, which is a function of effective temperature and luminosity. Furthermore, since we known the rotational frequency, $f_1=f_{\rm rot}=0.24730$ d$^{-1}$, the rotational velocity, $V_{\rm rot}$, can be derived. In the observed error box, the radius changes from 4.3 to 5.8\,R$_{\sun}$ and the rotational velocity from 54 to 72\,km\,s$^{-1}$. As a result, we have adopted $R$\,=\,$5.1\pm0.7$\,\rm{R}$_{\sun}$ and $V_{\rm {rot}}$\,=\,$63\pm9$\,km\,s$^{-1}$. Subsequently, with the value of $\vsini$\,=\,$40\pm3$ km\,s$^{-1}$, we were able to constrain the inclination angle to $i\in\,\langle31\deg$,\,53$\,\deg\rangle$. The mass of the star cannot be reliably determined from the HRD either. Its value depends strongly on the assumed model parameters as well as on the stellar evolution phase. However, the theoretically derived gravity value agrees well with the one derived from spectroscopy.

Fig.\,\ref{fig:Kiel} shows the Kiel diagram ($\log{T_{\rm{eff}}}$ vs. $\log{g}$) including the models fitting the observed values of $\log{T_{\rm {eff}}}$, $\log{L/L_{\sun}}$ and $f_{\rm {rot}}$, whose parameters are listed in Table\,\ref{tab:models}. The majority of models falls into the observed error box. Nevertheless, some systematics are noticeable, with the models tending to have smaller $\log{g}$ values than the central value of the error box. The models above the error box correspond to models during the advanced AtL stage of evolution.

\subsubsection{Pulsational models}

\hd\ is situated within the SPB instability strip \citep{1999AcA....49..119P,2017MNRAS.469...13S}; therefore, the occurrence of g mode pulsations can be expected. In this section, we therefore explore whether the secondary variability may have a pulsational origin.

Theoretical frequencies were calculated with the customised non-adiabatic pulsational code of \cite{1977AcA....27...95D,Dziembowski1977}, which solves linearised equations of stellar oscillations. Fig.\,\ref{fig:nueta} illustrates the instability parameter, $\eta$ \citep{Stellingwerf1978}, as a function of frequency. If $\eta>0$, the corresponding mode is excited in a model. We chose the main-sequence model of \hd\ calculated with $X_{\rm {ini}}=0.70$, $Z=0.015$ and $f_{\rm {ov}}=0.01$, which is situated in the center of the error box in the HRD. We considered modes of degrees $\ell=0,1,2$, and 3; for the sake of clarity, only centroid modes (modes with azimuthal number $m=0$) are shown in the plots. The vertical line in Fig.\,\ref{fig:nueta} marks the position of the observed frequency $f_2=0.47356$ d$^{-1}$. It becomes obvious that the observed frequency value $f_2$ is well within the predicted instability range of the model; in fact, frequencies of all considered mode degrees are excited near $f_2$.

Interestingly, our calculations indicate that instability in the low-frequency domain persists even after the main-sequence phase during the CoP (Fig.\,\ref{fig:nueta}, middle panel). However, the more advanced evolutionary phases are characterised by rather poor excitation of all modes (Fig.\,\ref{fig:nueta}, lower panel).

In summary, our models demonstrate that g mode pulsation frequencies, which are typically observed in SPB-type pulsators, are excited around the observed frequency $f_2$. It is therefore plausible to assume that the secondary light curve component is associated with g mode pulsation.

\begin{figure}
	\includegraphics[width=0.47\textwidth]{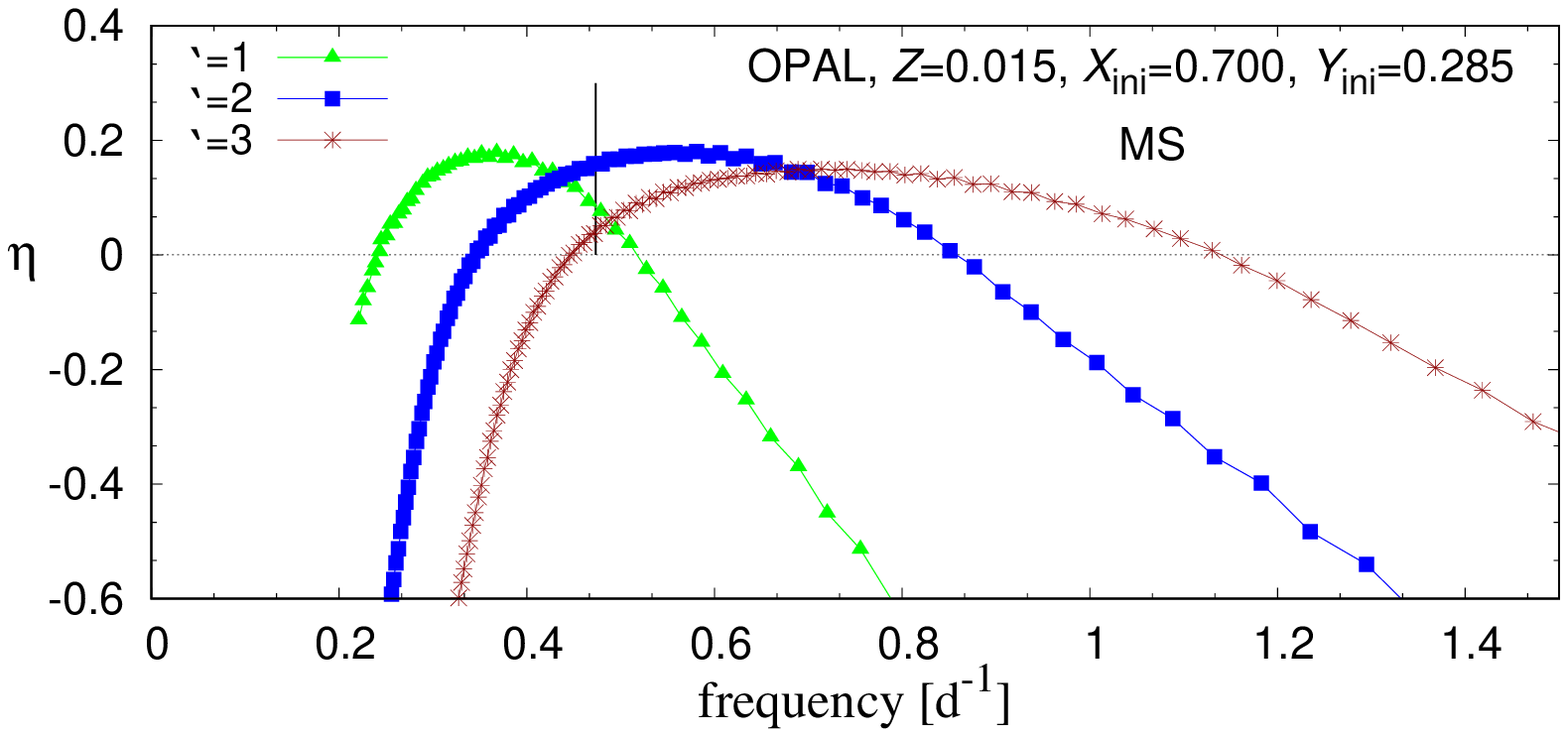}
	\includegraphics[width=0.47\textwidth]{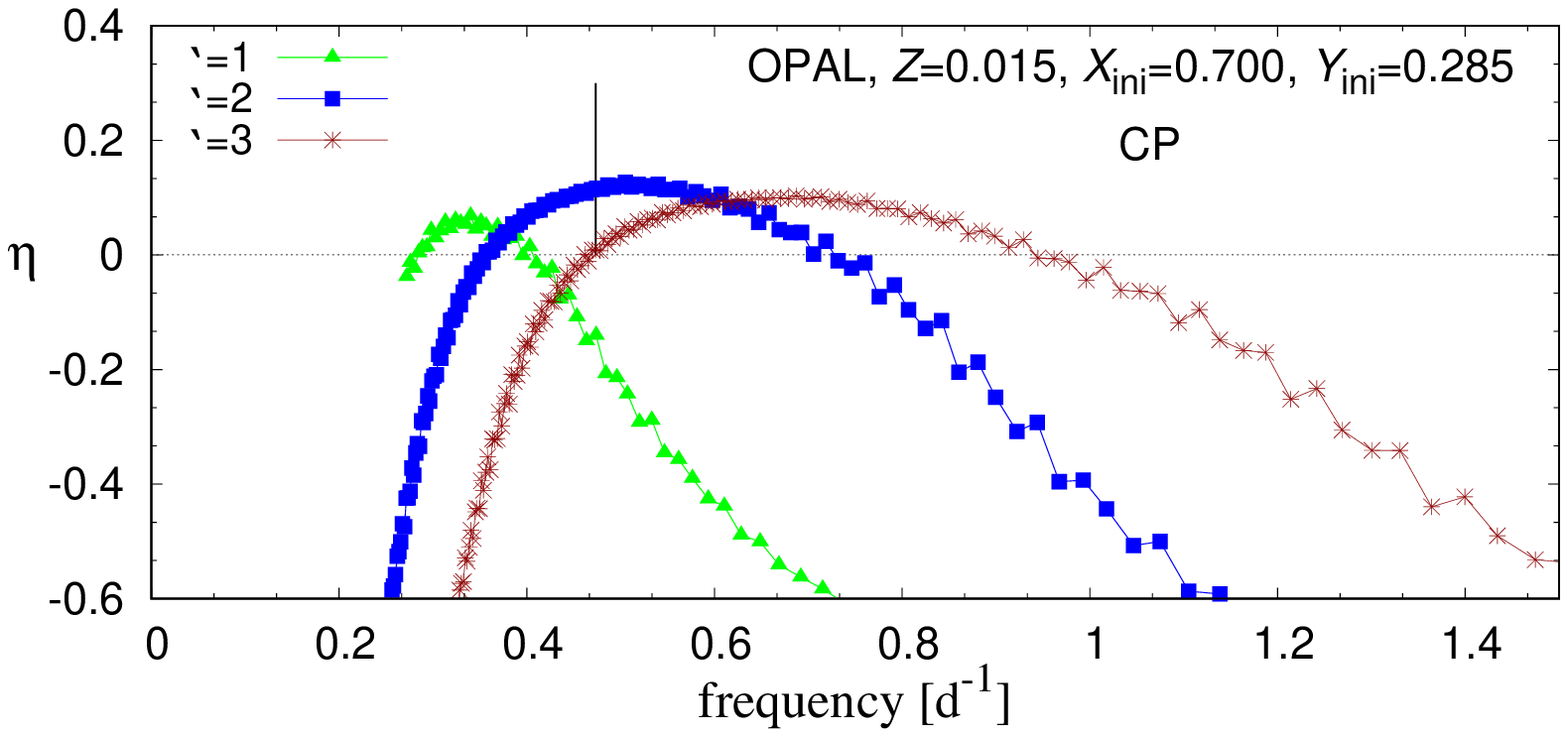}
	\includegraphics[width=0.47\textwidth]{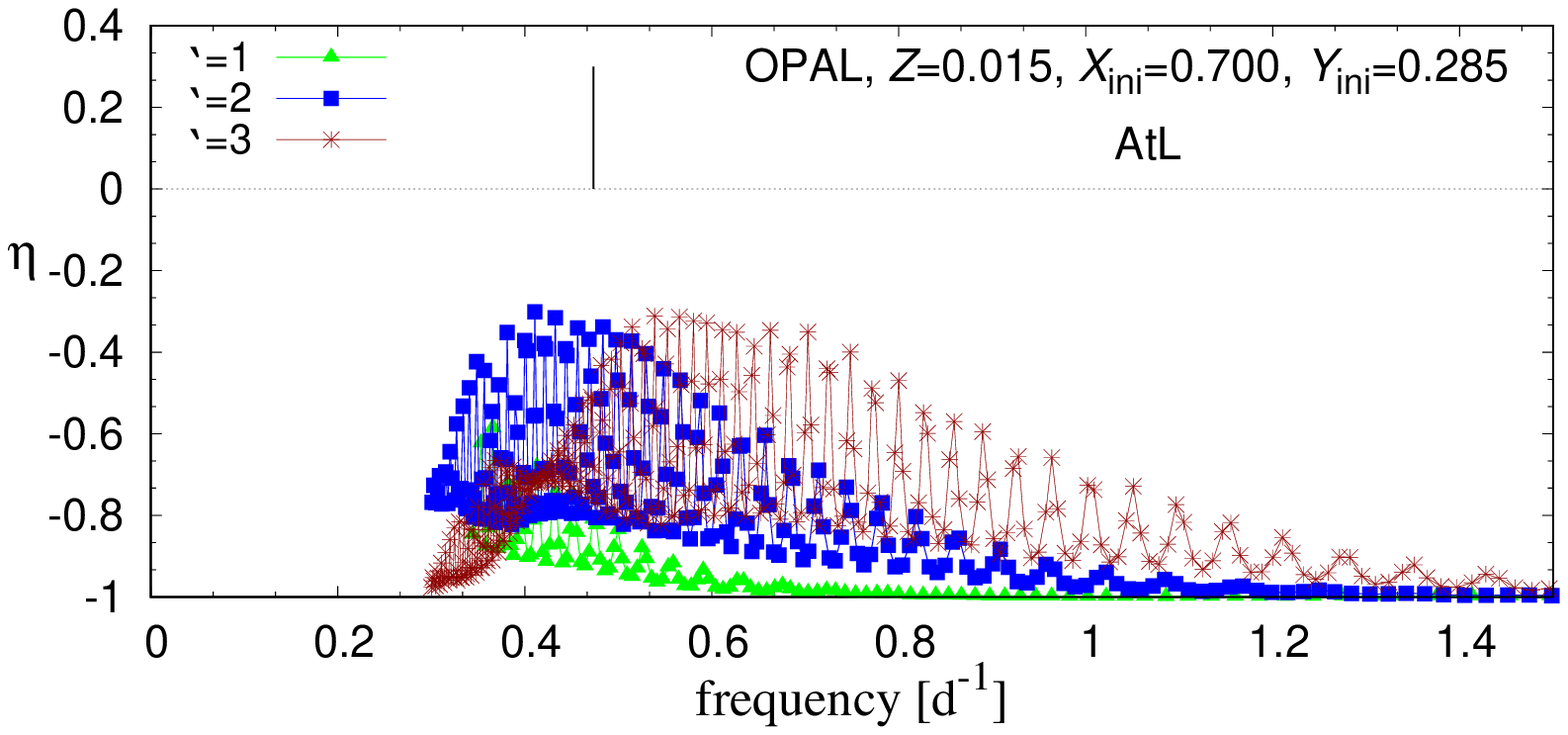}
	\caption{Instability parameter, $\eta$, as a function of frequency for the main sequence (upper panel), CoP (middle panel) and AtL (lower panel) models calculated with $Z=0.015$, $X_{\rm {ini}}=0.70$, $f_{\rm {ov}}=0.01$, $\log{T_{\rm {eff}}}=4.1206$ and $\log{L/L_{\sun}}=2.834$. Modes of degrees $\ell=1,2$ and 3 were considered.}
	\label{fig:nueta}
\end{figure}

\section{Conclusions}

Inspecting photometric observations obtained by the {\it Kepler}\,K2 survey, the B9pSi star \hd\ was found to exhibit an unusual light curve that can be interpreted as the sum of two independent strictly periodic signals with incommensurable periods. Replenishing K2 measurements with $V$ photometry from the ground-based ASAS-3 and ASAS-SN surveys, precise light curves of both components were derived. The corresponding periods were determined as $P_1=4.043\,61(5)$\,d and $P_2=2.111\,65(3)$\,d, with effective amplitudes of $B_1=17.14(3)$ and $B_2=8.77(4)$\,mmag. A careful investigation of the stars in the close angular vicinity of our target star indicated that they do not contribute in any significant way to the observed light variations.

Follow-up detailed spectroscopy confirmed that \hd\ is an mCP star. However, our observations do not indicate the presence of a longitudinal magnetic field stronger than 110\,G. The light variations of the principal component can be well interpreted by a simple rotator model with two or more large persistent photometric spots, which is in line with an mCP star classification. The source of the photometric variability of the secondary light curve component, however, is more difficult to explain.

Investigating whether \hd\ may be a binary system, we find that the observations are best fit by assuming a short period pulsating variable star companion with an effective amplitude larger than 0.15\,mag, whose light curve and period would be in line with a classical Cepheid pulsating in the first overtone. However, this very young object would have to be situated at a height of almost 3\,kpc below the Galactic disk and beyond the Galactic center. Such a configuration is very unlikely. Furthermore, none of the hypotheses based on the premise that \hd\ is a pair of unresolved periodically variable stars is able to justify the observed modulation of $\Delta RV$ with period $P_2$ (Fig.\,\ref{rvs}, panel 2c). In all studied spectra, our target star appears to be single-lined with no indications of a possible companion star. Apparently, therefore, the source of the variability of period $P_2$ originates in the single mCP star rotating with period $P_1$.

Consequently, after refusing the hypothesis of a single mCP star with strong latitudinal differential rotation, we further investigated the possibility that the variability of the secondary light curve component is due to g mode pulsation. In general, it is thought that chemically peculiarity and classical pulsations mutually exclude each other. However, \hd\ is located within the SPB instability strip, and our state-of-the-art pulsation models indicate the occurrence of g mode pulsations at the observed frequency. The secondary variability observed in \hd, therefore, is fully compatible with pulsation, although only a single pulsation mode has been distinguished from the available data.

In summary, based on all available data and in absence of a more plausible explanation, we put forth the hypothesis that the peculiar light variability of \hd\ is caused by rotational modulation due to the surface abundance patches of a Si-type mCP star and single-mode SPB-type pulsation. \hd, therefore, is a very rare object worthy of further detailed investigations. Although many efforts have been put into analyzing the variability of mCP stars \citep[e.g.][]{2018A&A...619A..98H,2019MNRAS.487.4695S,2020MNRAS.tmp..437B}, we are not aware of the existence of a similar object.

\hd\ may prove to be an excellent testbed for the investigation of the complex atmospheric phenomena caused by the interplay of the magnetic field, pulsation and chemical peculiarities.

\section*{Acknowledgements}

This work has been supported by the projects: DAAD (project No. 57442043), GA\,\v{C}R 18-05665S, and VEGA 2/0031/18 of the Slovak Academy of Sciences. The study was based on observations collected at the European Organisation for Astronomical Research in the Southern Hemisphere under ESO programmes 099.C-0081(A) and 099.A-9039(C).
LF acknowledges financial support from CNPq.
PW acknowledges support from the Polish National Science Centre grants 2015/17/B/ST9/02082 and 2018/29/B/ST9/01940. Calculations have been partly carried out using resources provided by the Wroclaw Centre for Networking and Supercomputing (http://www.wcss.pl), grant No. 265. 
MS acknowledges the financial support of the Operational Program Research,
Development and Education -- Project Postdoc@MUNI (No. CZ.02.2.69/0.0/0.0/16\_027/0008360).
TP would like to thank to the GINOP 2.3.2-15-2016-00003 of the Hungarian
National Research, Development and Innovation Office.
This work has also made use of data from the European Space Agency (ESA) mission {\it Gaia},
%(\url{https://www.cosmos.esa.int/gaia})
processed by the {\it Gaia} Data Processing and Analysis Consortium (DPAC) \url{https://www.cosmos.esa.int/web/gaia/dpac/consortium}).
Funding for the DPAC has been provided by national institutions, in particular the institutions participating in the {\it Gaia} Multilateral Agreement.
Based in part on observations obtained at the Southern Astrophysical Research (SOAR) telescope, which is a joint project of the Minist\'{e}rio da Ci\^{e}ncia, Tecnologia,
Inova\c{c}\~{o}es e Comunica\c{c}\~{o}es (MCTIC) do Brasil, the US National Science Foundation’s NSF’s NOIRLab (NOIRLab), the University of North Carolina at Chapel Hill (UNC), and Michigan State University (MSU).

\section*{Data availability}

The data underlying this article will be shared on reasonable request to the corresponding author.

\bibliographystyle{mnras}
\bibliography{HD174356}

\newpage

\appendix

\section{Spectroscopic data}\label{specdata}

\begin{figure*}
\begin{center}
\includegraphics[width=0.95\textwidth]{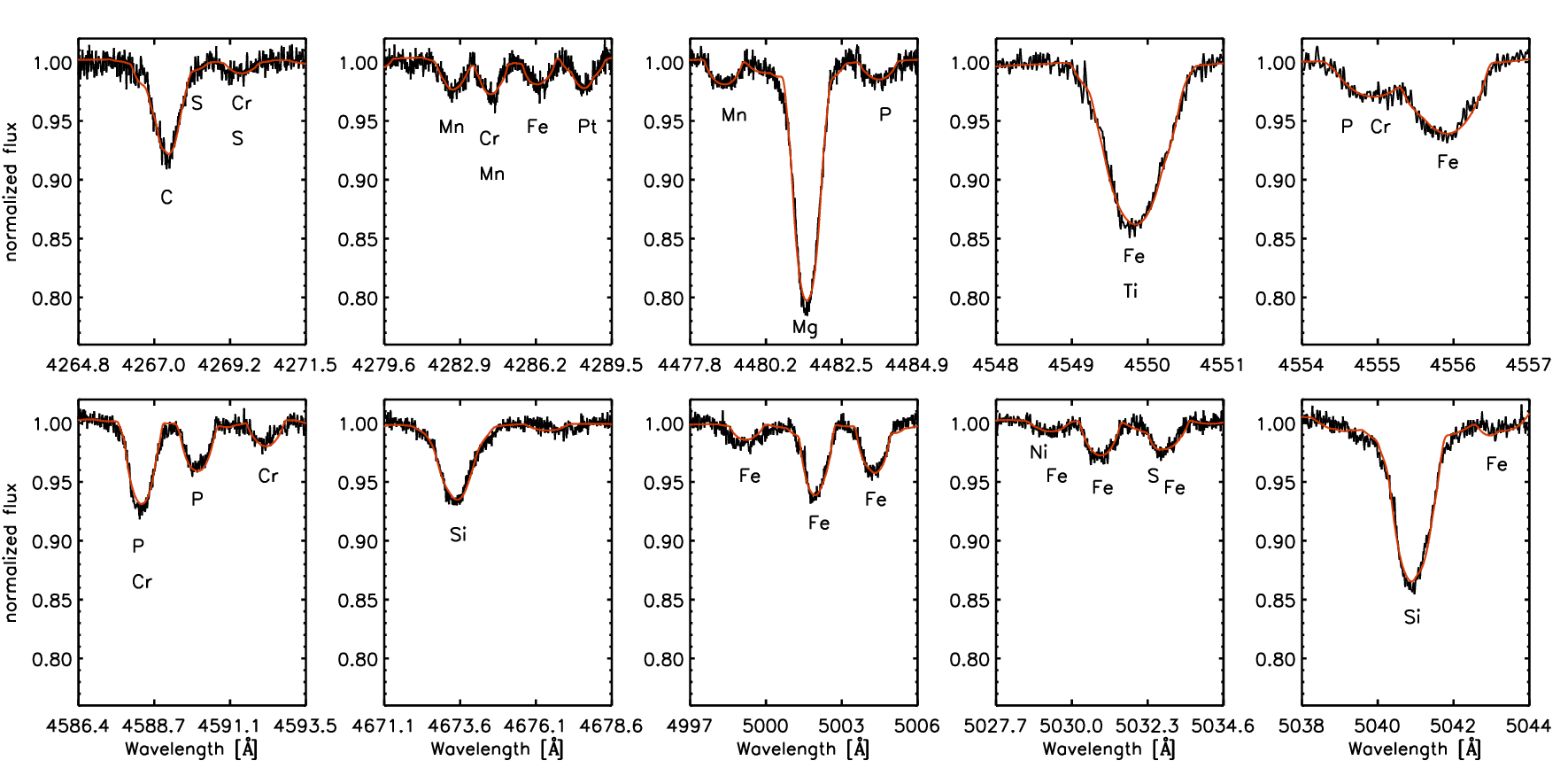}
\caption{Comparison of the observed (black) with the synthetic spectrum (pink) of \hd\ in selected spectral regions. The synthetic spectrum was calculated for \Teff = 13\,200\,K, $\log g = 3.8$, $\xi =0.5$\kms, and $v\sin i = 40$\,\kms.}
\label{lines}
\end{center}
\end{figure*}

\begin{table*}\scriptsize
\caption{Equivalent widths (in nm) and radial velocity (in \kms) of the \ion{Si}{II} \lam5041\,\AA\ and \lam5056\,\AA\ lines. Phase$_1$ and Phase$_2$ were calculated according to the ephemeris of the first light curve component (see Table\,\ref{tab}). The last two columns contain the measured radial velocity in relation to the mean value of $RV = +12.5(6)$\kms\ and the corresponding uncertainty.}
\label{SiEWRV}
\centering
\begin{tabular}{lccc|cc|cc|cc}
\hline\hline
&&&&\ion{Si}{II}&\ion{Si}{II}&\ion{Si}{II} &\ion{Si}{II}&$\Delta RV$&$\delta\Delta RV$ \\
Instrument &HJD\,+&Phase$_1$&Phase$_2$&\lam5041&\lam5056&\lam5041&\lam5056&All lines&All lines\\
&2450000&&&[nm]&[nm]& [\kms]&[\kms]&[\kms]&[\kms]\\
\hline
PST2    &    7839.948   & 0.299   & 0.013   &  0.147  &   0.127  &   13(4)     &    20(4)     &  +0.9 &	1.3   \\
PST2    &    7839.970   & 0.305   & 0.023   &  0.110  &   0.229  &   8(3)      &    10(3)     & $-$2.6 & 0.7 \\
PST2    &    7839.993   &  0.310   & 0.034   &  0.153  &   0.089  &   8(3)      &    4(3)      &  +2.0 &  1.0    \\
HIDES   &    7891.262   & 0.989   & 0.313   &  0.185  &   0.246  &   9.5(1.1)  &    20.4(1.0) &  +0.7 &  0.9    \\
HIDES   &    7892.255   & 0.235   & 0.784   &  0.185  &   0.269  &   10.6(1.0) &    18.6(9)   & $-$1.1 & 0.6  \\
HIDES   &    7893.246   &  0.480   & 0.253   &  0.174  &   0.222  &   10.0(1.1) &    19.2(1.0) & $-$0.5 & 0.5  \\
HIDES   &    7894.251   & 0.729   & 0.729   &  0.194  &   0.297  &   11.1(1.2) &    18.0(9)   & $-$0.2 & 0.7  \\
HARPS   &    7908.805   & 0.328   & 0.621   &  0.152  &   0.207  &   10.17(24) &    18.57(16) &  +2.5 &  0.6    \\
HARPS   &    7909.755   & 0.563   & 0.071   &  0.167  &   0.235  &   10.06(21) &    18.63(14) & $-$2.5 &  0.6    \\
HARPS   &    7910.738   & 0.806   & 0.536   &  0.157  &   0.220  &   9.34(25)  &    16.97(15) &  +1.1 &  0.6    \\
FEROS   &    7993.652   & 0.311   & 0.801   &  0.163  &   0.213  &   6.7(6)    &    14.8(3)   & $-$1.5 & 0.7  \\
FEROS   &    7994.665   & 0.561   & 0.281   &  0.177  &   0.228  &   9.0(6)    &    15.5(3)   &  +1.1 &  0.8    \\
FEROS   &    7996.659   & 0.054   & 0.225   &  0.179  &   0.234  &   9.2(8)    &    16.3(6)   &  +1.5 &  0.7    \\
CASLEO  &    8384.524   & 0.974   & 0.904   &         &          &             &              & $-$0.9 &  2.3\\
CASLEO  &    8385.501   & 0.217   & 0.367   &         &          &             &              &  +0.4 &  2.8\\
SOAR    &    8387.473   & 0.704   & 0.301   &  0.182  &   0.259  &   4.8(8)    &    13.7(8)   & $-$0.4 &  1.2  \\
SOAR    &    8387.477   & 0.705   & 0.303   &  0.186  &   0.250  &   9.1(9)    &    15.6(9)   &  +0.3 &  1.5 \\
SOAR    &    8387.481   & 0.706   & 0.304   &  0.191  &   0.263  &   13.8(1.1) &    18.6(8)   & $-$0.6 &  1.7  \\
SOAR    &    8389.506   & 0.207   & 0.263   &  0.176  &   0.236  &   13.8(1.1) &    21.2(7)   & $-$0.2 &  1.3 \\
SOAR    &    8389.509   & 0.208   & 0.265   &  0.158  &   0.224  &   1(6)      &    15(4)     & $-$0.5 & 1.4  \\
SOAR    &    8574.861   & 0.046   & 0.041   &  0.154  &   0.178  &   -0.8(1.2) &    8.1(9)    & $-$1.2 &  1.4 \\
SOAR    &    8574.868   & 0.048   & 0.044   &  0.164  &   0.198  &   2.7(1.0)  &    11.9(7)   & $-$0.9 &  1.3 \\
SOAR    &    8574.876   &  0.050   & 0.048   &  0.160  &   0.187  &   4.5(1.0)  &    10.3(7)   & $-$1.4 &  1.7  \\
SOAR    &    8574.890   & 0.053   & 0.054   &  0.159  &   0.249  &   9.7(1.1)  &    16.5(7)   & $-$0.6 &  2.0  \\
SOAR    &    8575.873   & 0.296   & 0.520   &  0.185  &   0.240  &   0.8(1.1)  &    10.2(8)   &  +0.5 &  1.5 \\
SOAR    &    8575.880   & 0.298   & 0.523   &  0.180  &   0.238  &   4.8(1.3)  &    13.5(9)   &  +0.2 &  2.1   \\
SOAR    &    8575.891   & 0.301   & 0.528   &  0.184  &   0.238  &   2.6(1.0)  &    10.7(8)   &  +0.6 & 1.6  \\
\hline
\end{tabular}
\end{table*}

\begin{table*}
\centering
\caption{
 Longitudinal magnetic field measurements using the LSD technique. All are non-detections
apart from the last measurement using metal lines showing a marginal detection.
}
\label{T:Bz}
\begin{tabular}{cccccc}
\hline
Date            & Phase            & $\langle B_z \rangle^{\mathrm{He}}$ &
FAP           & $\langle B_z \rangle^{\mathrm{met}}$ & FAP            \\
\hline
2017-06-04 & 0.328 & +144(140) & 0.708 & $-$212(69) & 0.872             \\
2017-06-05 & 0.563 & $-$351(135) & 0.996 & +16(60) & 0.002             \\
2017-06-06 & 0.806 & +247(95)  & 0.082 & $-$72(39) & $8\times 10^{-4}$ \\
\hline
\end{tabular}
\end{table*}

\begin{table*}
\scriptsize
\caption{Parameters of the $\ion{Si}{II}$ line equivalent width $EW$, radial velocity $RV$, and differential radial velocity $\Delta$ {\it RV} modeling (see Sect.\,\ref{rvs}). Radial velocity parameters are given in \kms, equivalent widths in nm. A discussion of the modeling is provided in Sect.\,\ref{rvs}.}
\label{spepar}
\centering
\begin{tabular}{cc|cccc|c}
\hline\hline
$\overline{EW}_1$ & $\overline{EW}_2$& $h_{11}$&$h_{12}$& $h_{21}$&$h_{22}$&$s_{EW}$\\
+0.168(5)&+0.226(7)&$-$0.014(27)&$-$0.037(34)&$-$0.034(28)&+0.019(37)&+0.022 nm\\
\hline
$\overline{RV}_1$ & $\overline{RV}_2$& $g_{11}$&$g_{12}$& $g_{21}$&$g_{22}$&$s_{RV}$\\
+18.2(6)&+10.2(7)&$-$0.3(6)&+0.8(6)&$-$0.4(6)&+1.8(8)&+1.6 \kms\\
\hline
$\overline{\Delta RV}$ && $a_{11}$&$a_{12}$& $a_{21}$&$a_{22}$&$s_{\Delta RV}$\\
$-$0.1(3)&&+0.4(4)&+0.3(4)&$-$1.5(6)&+0.3(3)&+1.2 \kms\\
\hline
\end{tabular}
\end{table*}

\clearpage

\section{Models of \hd}\label{pulsator}

\begin{figure}
\begin{center}
	\includegraphics[width=0.47\textwidth]{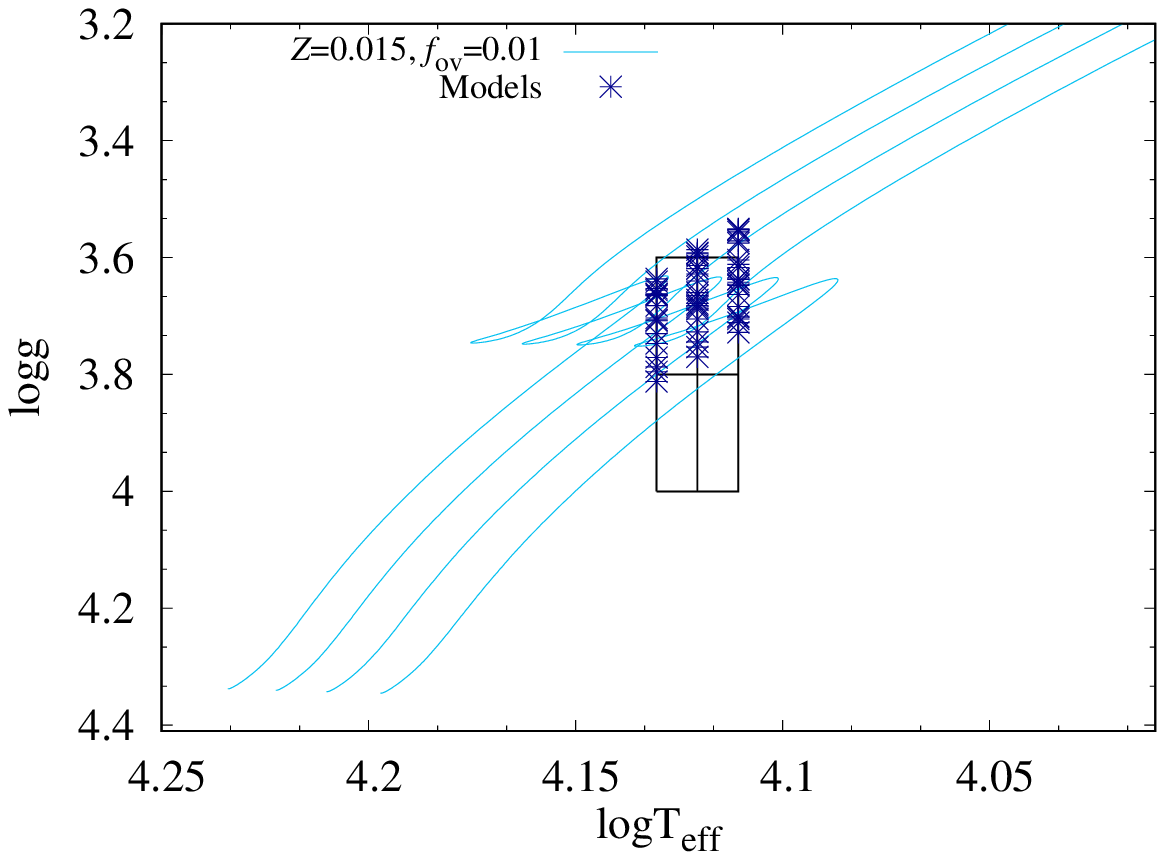}
	\caption{Kiel diagram with the position of HD\,174356 according to the parameters derived from spectroscopic analysis. The lines are evolutionary tracks for the given masses. The stars indicate the positions of the models listed in Table\,\ref{tab:models}.}
	\label{fig:Kiel}
\end{center}
\end{figure}

Table \ref{tab:models} is organised as follows:
\begin{itemize}
		\item Column 1: evolutionary phase of the model.
		\item Column 2: exponential overshooting parameter, $f_{\rm {ov}}$.
		\item Column 3: initial abundances of hydrogen, $X_{\rm{ini}}$.
		\item Column 4: initial abundances of helium, $Y_{\rm{ini}}$.
		\item Column 5: metallicity, $Z$.
		\item Column 6: stellar mass, $M$ [M$_{\sun}$].
		\item Column 7: age, calculated from cloud collapse [Myr].
		\item Column 8: effective temperature.
		\item Column 9: luminosity.
		\item Column 10: radius.
		\item Column 11: surface gravity.
		\item Column 12: central hydrogen content, $X_c$.
		\item Column 13: surface rotational velocity, $V_{\rm {rot}}$ [km\,s$^{-1}$].	
\end{itemize}

\begin{table*}\scriptsize
	\begin{center}
		\caption{Models of \hd\ that fit the observed values of effective temperature, luminosity and rotational frequency.}
		\label{tab:models}
		\begin{tabular}{cccccccccccccc}
			\hline
			\hline
		 Evol. Phase & $f{\rm {ov}}$ & $X_{\rm {ini} }$ & $Y_{\rm {ini}}$  & $Z$ & $M$ & age &  $\log{T_{\rm {eff}}}$ & $\log{L/L_{\sun}}$ & $R $ & $\log{g}$ & $X_{\rm{c}}$ & $V_{\rm {rot}}$   \\
			& & & & &$[\rm{M}_{\sun}]$ & [Myr]  &  &&$[R_{\sun}]$ & & & $[$km\,s$^{-1}$]\\		
			\hline	

MS & 0.01 & 0.7 & 0.285 & 0.015 & 4.2954 & 111.618 & 4.1304 & 2.7507 & 4.3421 & 3.7955 & 0.1913 & 54.35  \\
MS & 0.01 & 0.7 & 0.285 & 0.015 & 4.4756 & 105.722 & 4.1304 & 2.8338 & 4.7783 & 3.7302 & 0.1415 & 59.81  \\
MS & 0.01 & 0.7 & 0.285 & 0.015 & 4.6643 & 99.587 & 4.1304 & 2.9169 & 5.2580 & 3.6650 & 0.0900 & 65.81  \\
MS & 0.01 & 0.7 & 0.285 & 0.015 & 4.2686 & 116.985 & 4.1206 & 2.7507 & 4.5437 & 3.7533 & 0.1582 & 56.88  \\
MS & 0.01 & 0.7 & 0.285 & 0.015 & 4.4489 & 110.208 & 4.1206 & 2.8338 & 5.0003 & 3.6882 & 0.1080 & 62.59  \\
MS & 0.01 & 0.7 & 0.285 & 0.015 & 4.2444 & 122.048 & 4.1107 & 2.7507 & 4.7559 & 3.7113 & 0.1250 & 59.53  \\
\\
CoP & 0.01 & 0.7 & 0.285 & 0.015 & 4.2398 & 132.385 & 4.1304 & 2.8338 & 4.7781 & 3.7067 & 0.0020 & 59.81 \\
CoP & 0.01 & 0.7 & 0.285 & 0.015 & 4.5359 & 111.943 & 4.1304 & 2.9169 & 5.2581 & 3.6529 & 0.0081 & 65.81 \\
CoP & 0.01 & 0.7 & 0.285 & 0.015 & 4.2909 & 128.383 & 4.1206 & 2.8338 & 5.0003 & 3.6724 & 0.0052 & 62.59 \\
CoP & 0.01 & 0.7 & 0.285 & 0.015 & 4.5940 & 107.533 & 4.1206 & 2.9169 & 5.5024 & 3.6190 & 0.0267 & 68.88  \\
CoP & 0.01 & 0.7 & 0.285 & 0.015 & 4.4194 & 115.009 & 4.1107 & 2.8338 & 5.2328 & 3.6458 & 0.0711 & 65.50  \\
CoP & 0.01 & 0.7 & 0.285 & 0.015 & 4.3365 & 124.658 & 4.1107 & 2.8338 & 5.2328 & 3.6376 & 0.0129 & 65.50  \\
\\
AtL & 0.01 & 0.7 & 0.285 & 0.015 & 4.2527 & 131.567 & 4.1304 & 2.8338 & 4.7772 & 3.7081 & 0.0000 & 59.85  \\
AtL & 0.01 & 0.7 & 0.285 & 0.015 & 4.4173 & 119.957 & 4.1304 & 2.9169 & 5.2581 & 3.6414 & 0.0000 & 65.81  \\
AtL & 0.01 & 0.7 & 0.285 & 0.015 & 4.2186 & 134.288 & 4.1206 & 2.8338 & 5.0017 & 3.6650 & 0.0000 & 62.58  \\
AtL & 0.01 & 0.7 & 0.285 & 0.015 & 4.3784 & 122.679 & 4.1206 & 2.9169 & 5.5026 & 3.5981 & 0.0000 & 68.90  \\
AtL & 0.01 & 0.7 & 0.285 & 0.015 & 4.3525 & 124.607 & 4.1107 & 2.9169 & 5.7585 & 3.5560 & 0.0000 & 72.11  \\

\hline

MS & 0.01 & 0.73 & 0.255 & 0.015 & 4.4641 & 115.587 & 4.1304 & 2.7507 & 4.3421 & 3.8122 & 0.2239 & 54.35  \\
MS & 0.01 & 0.73 & 0.255 & 0.015 & 4.6520 & 109.649 & 4.1304 & 2.8338 & 4.7780 & 3.7470 & 0.1735 & 59.80 \\
MS & 0.01 & 0.73 & 0.255 & 0.015 & 4.8523 & 103.060 & 4.1304 & 2.9169 & 5.2581 & 3.6822 & 0.1244 & 65.81  \\
MS & 0.01 & 0.73 & 0.255 & 0.015 & 4.4364 & 121.321 & 4.1206 & 2.7507 & 4.5440 & 3.7700 & 0.1901 & 56.88  \\
MS & 0.01 & 0.73 & 0.255 & 0.015 & 4.6263 & 114.228 & 4.1206 & 2.8338 & 5.0004 & 3.7051 & 0.1408 & 62.59  \\
MS & 0.01 & 0.73 & 0.255 & 0.015 & 4.8240 & 107.087 & 4.1206 & 2.9169 & 5.5027 & 3.6402 & 0.0901 & 68.86 \\
MS & 0.01 & 0.73 & 0.255 & 0.015 & 4.4118 & 126.589 & 4.1107 & 2.7507 & 4.7554 & 3.7281 & 0.1573 & 59.52 \\
MS & 0.01 & 0.73 & 0.255 & 0.015 & 4.6017 & 118.613 & 4.1107 & 2.8338 & 5.2328 & 3.6633 & 0.1078 & 65.50  \\
\\
CoP & 0.01 & 0.73 & 0.255 & 0.015 & 4.1389 & 163.798 & 4.1107 & 2.7507 & 4.7553 & 3.7004 & 0.0010 & 59.52  \\
CoP & 0.01 & 0.73 & 0.255 & 0.015 & 4.4414 & 137.327 & 4.1107 & 2.8338 & 5.2334 & 3.6479 & 0.0050 & 65.51  \\
CoP & 0.01 & 0.73 & 0.255 & 0.015 & 4.3877 & 141.635 & 4.1206 & 2.8338 & 5.0005 & 3.6821 & 0.0019 & 62.59 \\
CoP & 0.01 & 0.73 & 0.255 & 0.015 & 4.6936 & 119.839 & 4.1206 & 2.9169 & 5.5026 & 3.6283 & 0.0077 & 68.87 \\
CoP & 0.01 & 0.73 & 0.255 & 0.015 & 4.6439 & 123.183 & 4.1304 & 2.9169 & 5.2576 & 3.6632 & 0.0032 & 65.81  \\
\\
AtL & 0.01 & 0.73 & 0.255 & 0.015 & 4.1860 & 159.339 & 4.1107 & 2.7507 & 4.7555 & 3.7053 & 0.0000 & 59.52  \\
AtL & 0.01 & 0.73 & 0.255 & 0.015 & 4.3838 & 142.203 & 4.1107 & 2.8338 & 5.2319 & 3.6423 & 0.0000 & 65.49  \\
AtL & 0.01 & 0.73 & 0.255 & 0.015 & 4.5524 & 129.752 & 4.1107 & 2.9169 & 5.7573 & 3.5756 & 0.0000 & 72.08  \\
AtL & 0.01 & 0.73 & 0.255 & 0.015 & 4.4140 & 139.740 & 4.1206 & 2.8338 & 4.9996 & 3.6847 & 0.0000 & 62.60  \\
AtL & 0.01 & 0.73 & 0.255 & 0.015 & 4.5926 & 126.891 & 4.1206 & 2.9169 & 5.5026 & 3.6188 & 0.0000 & 68.86  \\
AtL & 0.01 & 0.73 & 0.255 & 0.015 & 4.6343 & 124.044 & 4.1304 & 2.9168 & 5.2565 & 3.6623 & 0.0000 & 65.78  \\

\hline

MS & 0.02 & 0.7 & 0.285 & 0.015 & 4.2140 & 124.052 & 4.1304 & 2.7507 & 4.3422 & 3.7872 & 0.2346 & 54.34  \\
MS & 0.02 & 0.7 & 0.285 & 0.015 & 4.5657 & 110.601 & 4.1304 & 2.9169 & 5.2580 & 3.6558 & 0.1512 & 65.81 \\
MS & 0.02 & 0.7 & 0.285 & 0.015 & 4.1830 & 130.318 & 4.1206 & 2.7507 & 4.5440 & 3.7445 & 0.2059 & 56.89  \\
MS & 0.02 & 0.7 & 0.285 & 0.015 & 4.3548 & 122.721 & 4.1206 & 2.8338 & 5.0003 & 3.6789 & 0.1649 & 62.59  \\
MS & 0.02 & 0.7 & 0.285 & 0.015 & 4.5385 & 114.612 & 4.1206 & 2.9169 & 5.5025 & 3.6137 & 0.1253 & 68.87  \\
MS & 0.02 & 0.7 & 0.285 & 0.015 & 4.1556 & 136.023 & 4.1107 & 2.7507 & 4.7554 & 3.7022 & 0.1786 & 59.59  \\
MS & 0.02 & 0.7 & 0.285 & 0.015 & 4.3290 & 127.323 & 4.1107 & 2.8338 & 5.2328 & 3.6368 & 0.1389 & 65.50  \\
MS & 0.02 & 0.7 & 0.285 & 0.015 & 4.5123 & 118.504 & 4.1107 & 2.9169 & 5.7580 & 3.5717 & 0.0993 & 72.07  \\
\\
CoP & 0.02 & 0.7 & 0.285 & 0.015 & 4.0160 & 167.176 & 4.1107 & 2.8338 & 5.2328 & 3.6042 & 0.0005 & 65.50  \\
CoP & 0.02 & 0.7 & 0.285 & 0.015 & 4.3167 & 139.687 & 4.1107 & 2.9169 & 5.7585 & 3.5524 & 0.0030 & 72.07 \\
CoP & 0.02 & 0.7 & 0.285 & 0.015 & 4.2615 & 144.241 & 4.1206 & 2.9169 & 5.5025 & 3.5863 & 0.0011 & 68.88  \\
\\
AtL & 0.02 & 0.7 & 0.285 & 0.015 & 4.0855 & 160.219 & 4.1107 & 2.8338 & 5.2328 & 3.6117 & 0.0000 & 65.51  \\
AtL & 0.02 & 0.7 & 0.285 & 0.015 & 4.3141 & 140.130 & 4.1107 & 2.9169 & 5.7580 & 3.5522 & 0.0000 & 72.05  \\
AtL & 0.02 & 0.7 & 0.285 & 0.015 & 4.3173 & 139.798 & 4.1206 & 2.9169 & 5.5019 & 3.5920 & 0.0000 & 68.88  \\
\hline
MS & 0.01 & 0.7 & 0.29 & 0.01 & 4.0943 & 119.097 & 4.1304 & 2.7507 & 4.3422 & 3.7747 & 0.1281 & 54.35  \\
MS & 0.01 & 0.7 & 0.29 & 0.01 & 4.0678 & 124.309 & 4.1206 & 2.7507 & 4.5440 & 3.7324 & 0.0922 & 56.87  \\
\\
CoP & 0.01 & 0.7 & 0.29 & 0.01 & 3.9166 & 143.624 & 4.1304 & 2.7507 & 4.3424 & 3.7554 & 0.0035 & 54.35  \\
CoP & 0.01 & 0.7 & 0.29 & 0.01 & 3.9573 & 139.820 & 4.1206 & 2.7507 & 4.5446 & 3.7204 & 0.0084 & 56.88 \\
CoP & 0.01 & 0.7 & 0.29 & 0.01 & 4.0084 & 134.090 & 4.1107 & 2.7507 & 4.7553 & 3.6865 & 0.0287 & 59.52 \\
\\
AtL & 0.01 & 0.7 & 0.29 & 0.01 & 3.8816 & 147.051 & 4.1304 & 2.7507 & 4.3408 & 3.7515 & 0.0000 & 54.38 \\
AtL & 0.01 & 0.7 & 0.29 & 0.01 & 4.0252 & 134.829 & 4.1304 & 2.8338 & 4.7782 & 3.6842 & 0.0000 & 59.81  \\
AtL & 0.01 & 0.7 & 0.29 & 0.01 & 4.1974 & 122.101 & 4.1304 & 2.9169 & 5.2582 & 3.6192 & 0.0000 & 65.81  \\
AtL & 0.01 & 0.7 & 0.29 & 0.01 & 3.8386 & 151.227 & 4.1206 & 2.7507 & 4.5442 & 3.7072 & 0.0000 & 56.87  \\
AtL & 0.01 & 0.7 & 0.29 & 0.01 & 3.9951 & 137.445 & 4.1206 & 2.8338 & 5.0008 & 3.6414 & 0.0000 & 62.59  \\
AtL & 0.01 & 0.7 & 0.29 & 0.01 & 4.1789 & 123.516 & 4.1206 & 2.9169 & 5.5025 & 3.5778 & 0.0000 & 68.87  \\
AtL & 0.01 & 0.7 & 0.29 & 0.01 & 3.8060 & 154.538 & 4.1107 & 2.7507 & 4.7560 & 3.6640 & 0.0000 & 59.50  \\
AtL & 0.01 & 0.7 & 0.29 & 0.01 & 3.9756 & 139.223 & 4.1107 & 2.8338 & 5.2336 & 3.5998 & 0.0000 & 65.49 \\
			\hline
			\hline
		\end{tabular}
	\end{center}
\end{table*}
\label{lastpage}
\end{document}